\documentclass{jfm}
\usepackage{graphicx}
\usepackage{amsmath,amssymb,amsbsy,epstopdf,subfigure}
\usepackage[a4paper, left=2cm, right=2cm, top=2cm, bottom=2cm]{geometry}
\usepackage{natbib}

\pdfoutput=1

\newcommand{\pdiff}[3][]{\dfrac{\partial^{#1} #2}{\partial {#3}^{#1}}}

\title[Coalescence in a viscous gas]
{A parametric study of the coalescence of liquid drops in a viscous gas}

\author[J.E. Sprittles and Y.D. Shikhmurzaev]
{J\ls A\ls M\ls E\ls S\ns E.\ns S\ls P\ls R\ls I\ls
T\ls T\ls L\ls E\ls S\footnote{E-mail: J.E.Sprittles@warwick.ac.uk} \and Y\ls U\ls L\ls I\ls I\ns D.\ns S\ls H\ls I\ls K\ls H\ls M\ls U\ls
R\ls Z\ls A\ls E\ls V}

\affiliation{Mathematics Institute, University of Warwick, Coventry, CV4 7AL, UK, \newline
School of Mathematics, University of Birmingham, Birmingham B15 2TT, UK.}

\begin{document}

\label{firstpage} \maketitle

\begin{abstract}
The coalescence of two liquid drops surrounded by a viscous gas is
considered in the framework of the conventional model. The problem is solved
numerically with particular attention to resolving the very initial
stage of the process which only recently has become accessible both
experimentally and computationally.   A systematic study of the parameter space of practical interest allows the influence of the governing parameters in the system to be identified and the role of viscous gas to be determined. In particular, it is shown that the viscosity of the gas suppresses the formation of toroidal bubbles predicted in some cases by early computations where the gas' dynamics was neglected.  Focussing computations on the very initial stages of coalescence and considering the large parameter space allows us to examine the accuracy and limits of applicability of various `scaling laws' proposed for different `regimes' and, in doing so, reveal certain inconsistencies in recent works.  A comparison to experimental data shows that the conventional model is able to reproduce many qualitative features of the initial stages of coalescence, such as a collapse of calculations onto a `master curve' but, quantitatively, overpredicts the observed speed of coalescence and there are no free parameters to improve the fit.  Finally, a phase diagram of parameter space, differing from previously published ones, is used to illustrate the key findings.
\end{abstract}

\section{Introduction}

\begin{figure}[h]
     \centering
     \begin{minipage}[l]{.99\textwidth}
\subfigure[$t_i=0.01$]{\includegraphics[scale=0.25]{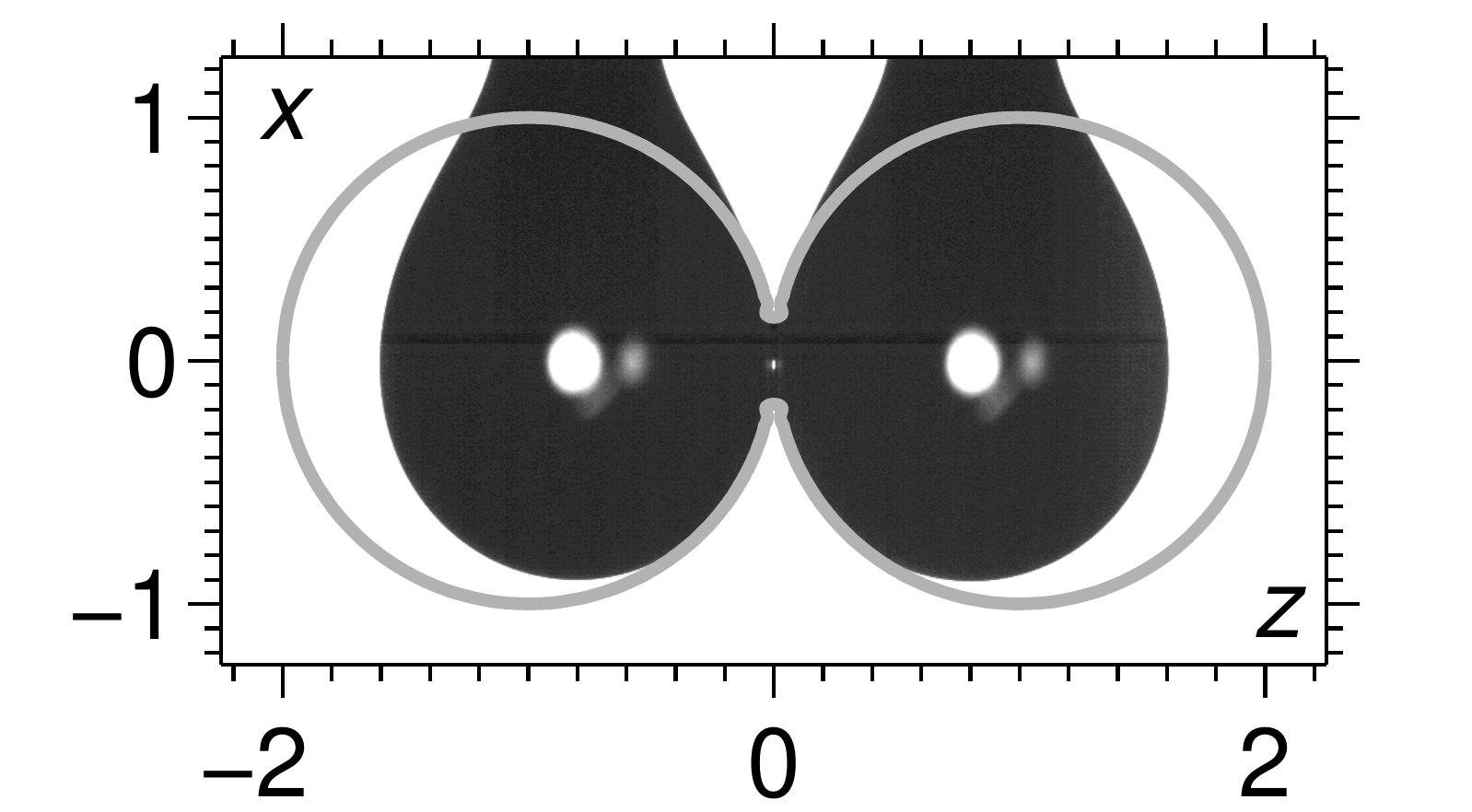}}
\subfigure[$t_i=0.07$]{\includegraphics[scale=0.25]{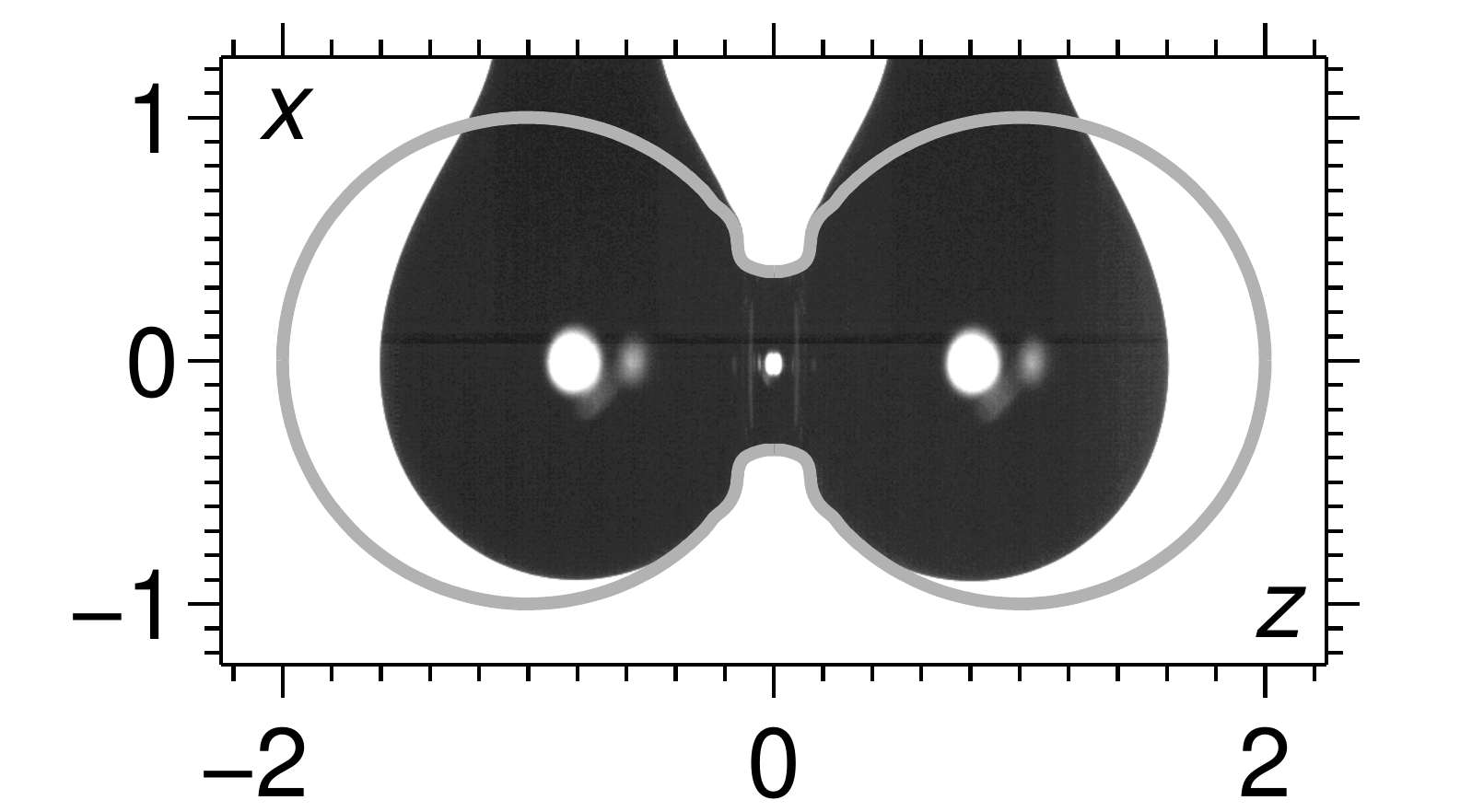}}
\subfigure[$t_i=0.15$]{\includegraphics[scale=0.25]{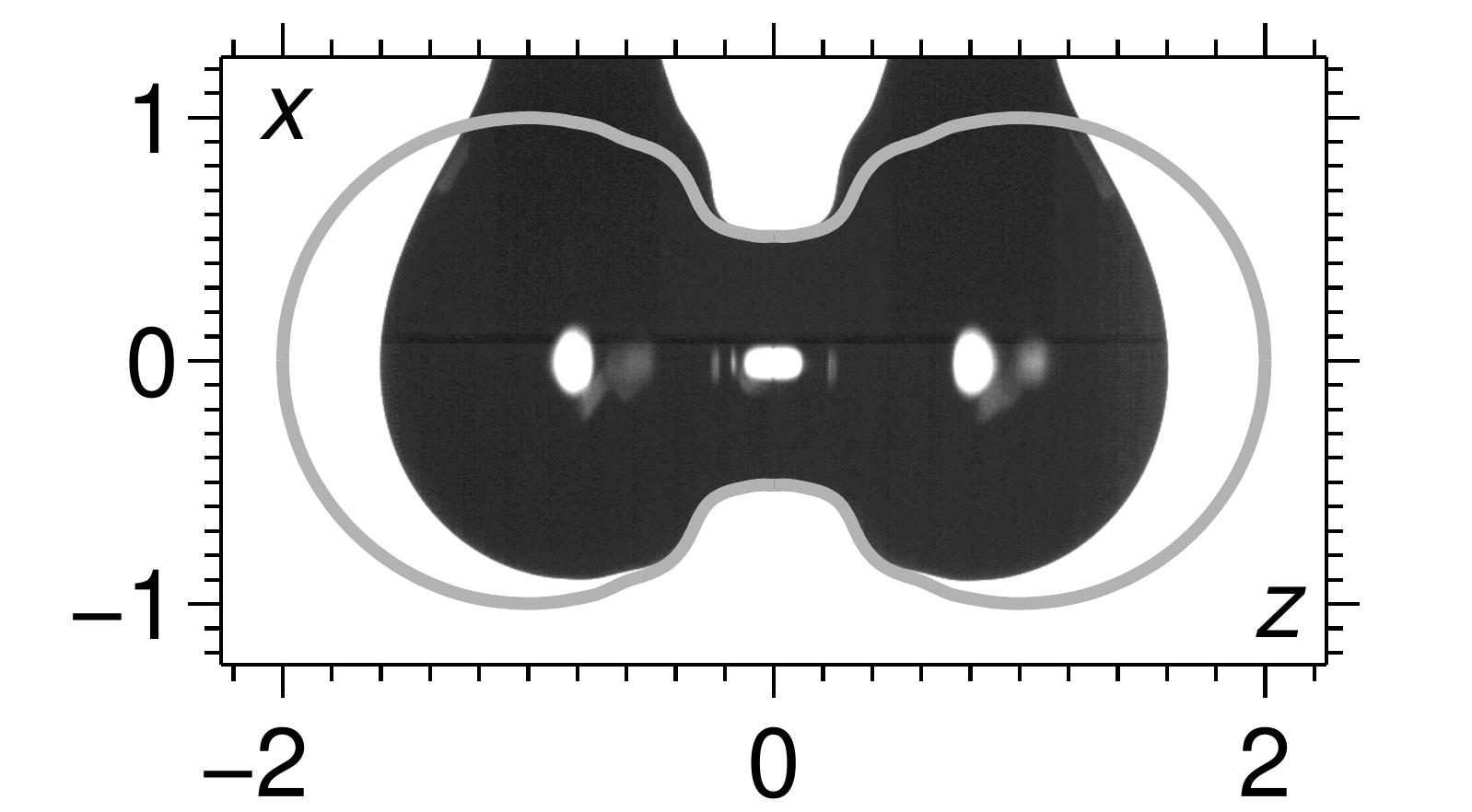}}
\subfigure[$t_i=0.3$]{\includegraphics[scale=0.25]{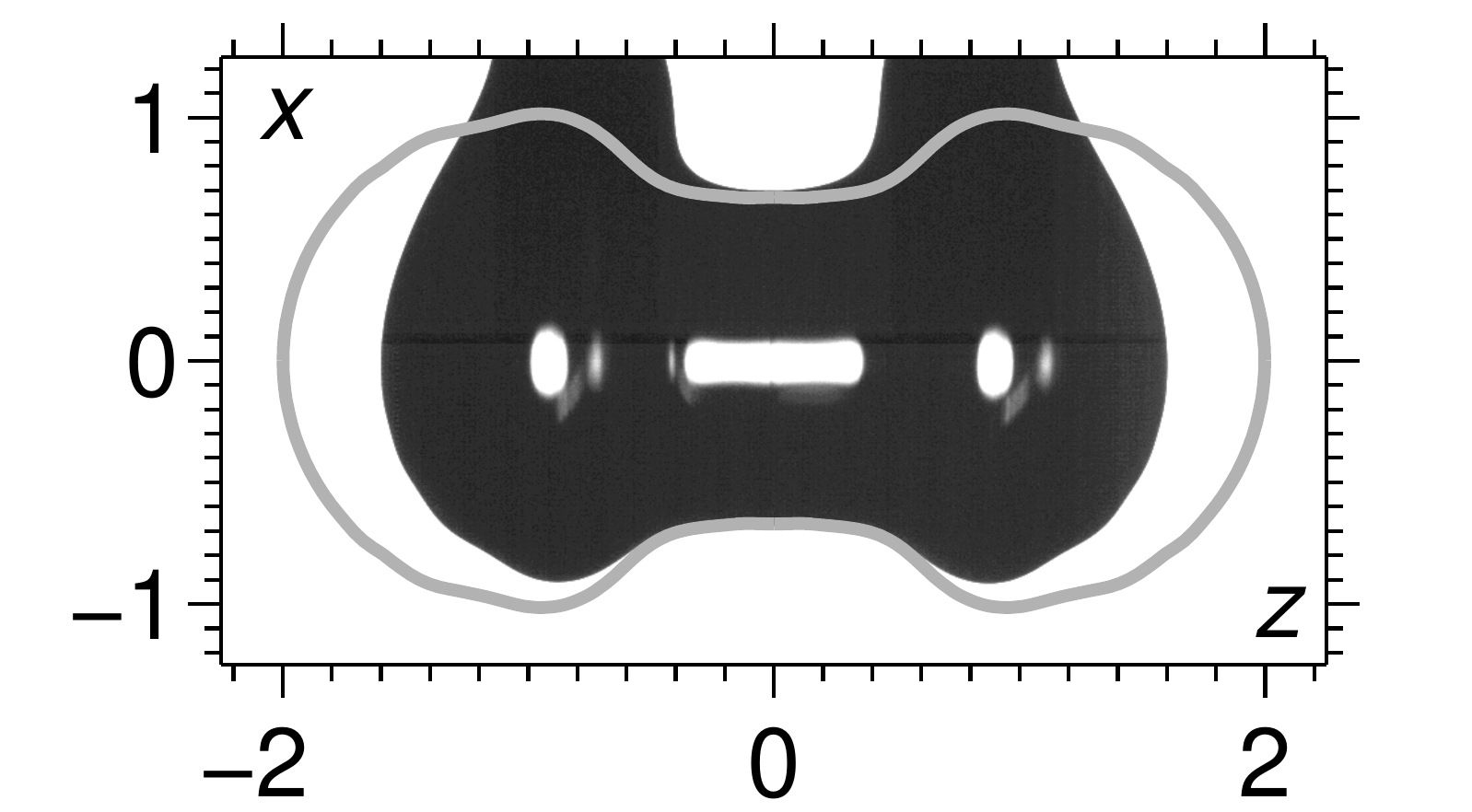}}
    \end{minipage}
 \caption{Comparison of our coalescence computations with free spheres in the inertial regime, against experiments in \cite{paulsen12} conducted using $1$~mm radii pendant drops of silicone oil with $Re=1.9\times 10^{4}$.  The dimensionless time $t_i$ is based on the inertial scale.  As can be seen, the neck region is accurately described far beyond the initial stages of the process even though the global geometry is different.}\label{F:inertial_pics}
\end{figure}

When two drops of the same liquid come into contact, a coalescence
process merges the two distinct bodies of liquid into one, after
which the resulting single body evolves towards its
equilibrium shape (Figure~\ref{F:inertial_pics}). This process can be observed in a range of
natural phenomena and holds the key to a vast number of emerging
technologies such as the `3D-printers' used to additively
manufacture complex products by assembling liquid microdrops in
`2D-slices' \citep{derby10} or the coalescence-induced jumping
mechanism being harnessed to enhance the heat transfer properties of
a solid covered by a condensed liquid \citep{enright12}.  Although
the equilibrium configuration of such systems is sometimes known,
the dynamics of the process that leads to it is not always well
understood. An example of unexpected dynamic behaviour is the
repeated partial coalescence of an ever decreasing sized drop with a
liquid bath, the so-called `coalescence cascade', observed by ultra
high-speed imaging techniques \citep{thoroddsen00}.

Although improving optical techniques have made it possible to study
small-scale high-speed free-surface flows \citep{thoroddsen08}, they
have intrinsic limitations associated with their spatial resolution
and, in particular, are unable to resolve the cusp-like region
formed when two drops are pressed into one another, or when one drop
is pressed into a solid \citep{eddi13}. As a result, one can often
only observe the appearance of the bridge between the two drops when
it has already travelled $\sim$10\% of the initial drop radius,
i.e.\ long after what one would class as the initial stages of
coalescence as such, where the merging of the two liquid bodies into
one has already occurred. An alternative technique, based on
measuring the electrical resistance of the bridge connecting the
drops, has been applied in \cite{paulsen11,paulsen14} where, for the first
time, the sub-micron scales of the coalescence phenomenon have been
resolved. This offers a unique opportunity to compare the predictions of the conventional model, i.e.\ simply the classical equations of hydrodynamics (incompressible Navier-Stokes equations with the surface tension of the liquid-gas interface assumed constant), which are known to be `singular' for this problem \citep{eggers99}, with the new experiments for the initial stages of the coalescence phenomenon at unprecedentedly small spatio-temporal scales.

In a recent publication \citep{sprittles_pof2}, the coalescence of
liquid drops in an inviscid dynamically-passive gas, henceforth referred to as a `passive gas', was computed, in
the framework of two different mathematical models, by adapting a
finite-element code initially developed for dynamic wetting
phenomena \citep{sprittles_ijnmf,sprittles_jcp,sprittles_pof}. The results were compared to experiments from both electrical
measurements in \cite{paulsen11} and optical ones in
\cite{thoroddsen05}. The first model examined was the conventional
one used in most studies, e.g.\ in \cite{eggers99}, and the one considered in this work. Its essence is
that, after the two drops touch at a point, it is assumed that an
infinitesimal smooth liquid bridge is formed that connects them, so
that the coalescence as such is actually over. The model is
concerned with the subsequent process, namely how the Laplacian
capillary pressure due to the highly curved free surface drives the
already formed single body of liquid towards its equilibrium shape.
The results of our numerical computations \citep{sprittles_pof2}
showed that the conventional model of the coalescence phenomenon,
whose solution is known to contain singularities in, amongst other
things, the radial velocity at the start of the process
\citep{hopper84,hopper90,hopper93a,hopper93b,richardson92},
overshoots experimental data from \cite{paulsen11}, i.e.\ it overpredicts the speed at
which coalescence occurs, whilst a singularity-free model, incorporating interface formation dynamics \citep{shik07}, captures the data more accurately.  This model has recently been the subject of further, more detailed, investigation in \cite{sprittles14_jfm1}.

Notably, in \cite{sprittles_pof2} the main emphasis was on a direct comparison between the two aforementioned models and experimental data.  In contrast, here our attention will be focussed entirely on the conventional model, which, so far, remains the most popular approach to describe such flows, with computations resolving both the fine-scales associated with the initial bridge propagation right through to the scales on which the overall dynamics of the coalescing drops comes into play.  In particular, we will report on the results of a full parametric study of the coalescence process which allowed us to (a) determine the role of parameters in the model; (b) identify different `regimes' proposed in the published literature and the crossovers between them; and (c) calculate the accuracy of `scalings' proposed for these regimes. As a result of the comprehensive comparison between our computations and previous theoretical works on the coalescence phenomenon we will identify a number of discrepancies in the previous published literature.

Furthermore, given that, as shown in \cite{sprittles_pof2}, the conventional model overpredicts the speed of coalescence compared to data from experiments, where the exterior fluid was air, i.e.\ a viscous gas, one could argue that the overshoot could, perhaps, be attributed to the neglect of the gas' dynamics.  For example, one may argue that the high pressures needed to squeeze the gas out of the cusp-like region at the bridge-front, if accounted for in the model, could slow the front down. Therefore, in the present paper, we also include {\it two-phase\/} calculations of the coalescence phenomenon into our parametric study and make a direct comparison of the results to the experimental data.

\section{Asymptotic results and `scaling laws' for the coalescence of liquid
drops}\label{S:asymp}

Simplified expressions for the coalescence event which are valid in different `regimes', have gained popularity due
to their simplicity compared to the full-scale theoretical description for, in particular, providing explicit formulas
to fit experimental data.  On the theoretical level, in the framework of the conventional model, the most commonly used results are those in \cite{hopper84}, where conformal mapping techniques have been used to derive an \emph{exact} solution to the problem of two-dimensional viscous-dominated coalescence. On the level of the scaling laws, the most frequently used ones were derived in \cite{eggers99}, for both viscous- and inertia-dominated coalescence. The recent results in \cite{paulsen12} suggest the existence of a third inertially-limited viscous regime which precedes all others.  The results of these works will be subject to scrutiny in the forthcoming sections, and are therefore now briefly described.

\subsection{Viscous-dominated regime}

The natural scale for velocity in this regime is given by
$U_{v}=\sigma/\mu$, where $\mu$ is the liquid's viscosity and $\sigma$ is the surface tension of the liquid-gas interface, so that
the capillary number $Ca=\mu U_{v}/\sigma=1$. The appropriate time
scale is then $T_{v}=R\mu/\sigma$, where $R$ is the drop's initial
radius, which is the characteristic length scale in all regimes. The Reynolds number then becomes $Re=\rho\sigma R/\mu^2$, where $\rho$ is the liquid's density. Alternatively, some works, e.g.\ \cite{paulsen12}, characterise the coalescence in terms of the Ohnesorge number which is related to the Reynolds number by $Oh=Re^{-1/2}$.  Henceforth, unless denoted by a subscript `dim' to denote `dimensional', all quantities will be assumed dimensionless.

\subsubsection{Exact solution in \cite{hopper84}}

The exact result obtained in \cite{hopper84} gives the entire two-dimensional shape of two identical coalescing cylinders, described by Stokes flow, in a passive gas as a function of time.  What will be of most interest to us in characterising the coalescence event are the bridge radius $r$ and height of the drops $h$ (Figure~\ref{F:sketch}) as a function of time $t$, which are given by
\begin{equation}\label{hopper}
r = \sqrt{2} (1-m)(1+m^2)^{-1/2}, \qquad h = \sqrt{2} (1+m)(1+m^2)^{-1/2}
\end{equation}
where the parameter $m$ is related to the time by
\begin{equation}\label{hopper1}
t = \frac{\pi\sqrt{2}}{4}\int^{1}_{m^2}\left[\tau(1+\tau)^{1/2}K(\tau)\right]^{-1}d\tau ,\qquad K(\tau) = \int_{0}^{1} \left[(1-x^2)(1-\tau x^2)\right]dx,
\end{equation}
an expression which can easily be evaluated numerically.

\subsubsection{Scaling law in \cite{eggers99}}

The scaling laws in \cite{eggers99} are derived by balancing the driving capillary pressure
$\sigma \kappa$, where $\kappa$ is the curvature at the bridge front, with the key resistive mechanism, i.e.\ either viscous or inertial forces.  In both cases, the driving force is shown to result primarily
from the longitudinal curvature so that $\kappa \propto 1/d(t)$, where
$d(t)$ is the longitudinal radius of curvature at the bridge front
(Figure~\ref{F:sketch}).

In the viscous-dominated case, it is shown that, local to the bridge front for $r\ll1$, the two-dimensional solution from \cite{hopper84} can be used to provide the radius of curvature, which scales like $r^\alpha$, where $\alpha=3$. In other words, it is assumed that the
evolution of 2D and 3D drops are identical in the initial stages. It is then further argued that $\alpha=3/2$ when the gas has some viscosity $\mu_g$.

As a result, the expression for the (dimensionless) bridge radius for $r\ll 1$ has the form
\begin{equation}\label{eggers}
r = -C_v t\ln t,\qquad C_v=\frac{\left(\alpha-1\right)}{2\pi}, \qquad   \alpha=\left\{
                                                                                  \begin{array}{ll}
                                                                                    3, & \hbox{$\bar{\mu}=0$;} \\
                                                                                    3/2, & \hbox{$\bar{\mu}>0$.}
                                                                                  \end{array}
                                                                                \right.
\end{equation}
Notably, and somewhat counter-intuitively, when the external fluid is regarded to be viscous,
the form of equation (\ref{eggers}) does not depend on the gas-to-liquid viscosity ratio $\bar{\mu}=\mu_g/\mu$, and it is only $\alpha$ which changes from $3$ to $3/2$, although, it is specified in \cite{eggers99} that the region of applicability of the formula should depend on this parameter; (\ref{eggers}) is expected to hold for $r<\bar{\mu}^{2/3}$.

\subsection{Inertia-dominated regime}

The characteristic scale for velocity in the inertia-dominated
regime is obtained by setting the Weber number to unity, so that
$U_{i}=\sqrt{\sigma/(\rho R)}$.  The characteristic time scale for this regime is then given
by $T_{i}=\sqrt{\rho R^3/\sigma}$. The Reynolds number in the inertia-dominated regime $Re_i$ is related to the one in the viscous regime $Re$ by $Re_i = Re^{1/2}$.

In \cite{eggers99}, it is suggested that the driving capillary pressure due to the surface
tension and based on the longitudinal curvature obtained from the
undisturbed free-surface shape of the drops $d(t)\sim r_{dim}^2(t_{dim})/R$ is
balanced by the dynamic pressure $\rho \left(dr_{dim}/dt_{dim}\right)^2$. As a
result, one has $r_{dim}/R = C_{i}\left(t_{dim}/T_{i}\right)^{1/2}$, where
$C_{i}$ is a constant of proportionality, so that, once
non-dimensionalised by our characteristic scales in this regime, the
scaling law takes the form
\begin{equation}\label{io}
r=C_{i}t_i^{1/2}.
\end{equation}
where $t_i$ is time made dimensionless by $T_i$.

Notably, in contrast to (\ref{eggers}), there is no closed-form expression for
$r(t)$, as the expression contains an unknown prefactor.  These issues are addressed in further detail in \cite{sprittles14_pre}.

\subsection{Inertially-limited viscous regime}

Recently, an `inertially-limited viscous' (ILV) regime has been shown in \cite{paulsen12}, through a combination of experimental and computational techniques, to precede either the viscosity-dominated regime or the inertia-dominated one, for non-zero values of $Re$, see also \cite{paulsen13,paulsen14}.  In particular, it is noted that in Hopper's exact solution (\ref{hopper}), for Stokes flow, once coalescence commences, the entire volume of each drop is translated towards the other, so that the motion cannot be considered as `local' to the neck region, as in (\ref{eggers}) and (\ref{io}).  Such global motion can be observed, for example, by measuring the height of the drops, i.e.\ a position far away from the bridge, as a function of time.

In \cite{paulsen12}, it is shown that for finite $Re$ the neck must reach a finite radius \emph{before} it has enough force to create this global motion; until it does so, it is in the ILV regime. Experiments suggest that in this regime the bridge propagates at a constant speed, which is simply determined from dimensional analysis to be $U_v$.  This gives
\begin{equation}\label{ilv}
r=C_{v}t.
\end{equation}
where, in contrast to (\ref{eggers}), $C_v$ is an a-priori unknown prefactor.

\section{Overview of the study}

Although many experimental and theoretical studies have considered the various regimes, and the crossovers between them, there has been no systematic parametric study of the system using a full-scale theoretical description accounting for viscous, inertial and capillary effects as well as the influence of the ambient fluid surrounding the coalescing drops.  Furthermore, computations have tended to either focus on only the very initial stages of the process, often using boundary integral methods to look only at the viscous regime \citep{eggers99} or the inertial one \citep{oguz89,duchemin03}, or on the global dynamics, with the initial stages not considered.  As a result, in none of these works have the influence of a viscous gas been considered in detail.  It is this gap in the theoretical research on coalescence which we shall now address and, as a by-product, uncover and examine various inconsistencies in the published literature.

In section~\ref{S:model}, the problem formulation is given for both the case of free spheres coalescing as well as the pinned hemispherical drop configuration often considered experimentally (Figure~\ref{F:sketch}). Section~\ref{S:comp} describes the main elements of our computational approach including, when required, references to more detailed expositions.  Results are presented in section~\ref{S:para}, where a full systematic study of parameter space is performed which elucidates, in particular, the effect of both the liquid's and the outer gas' properties.  At each stage, a detailed comparison with the previous literature, summarised in section~\ref{S:asymp}, is provided.  The full parametric study is followed by a comparison to experimental results both from qualitative and quantitative perspective in section~\ref{S:exp}.  The results from sections~\ref{S:para} and \ref{S:exp} are tied in with the published literature in section~\ref{S:disc}, where it proves illustrative to represent our findings with a phase diagram.  Final conclusions and, motivated by our results, suggestions for new directions of experimental and theoretical research are given in section~\ref{S:conc}.

\section{Problem formulation}\label{S:model}

\begin{figure}
     \centering
\includegraphics[scale=0.65]{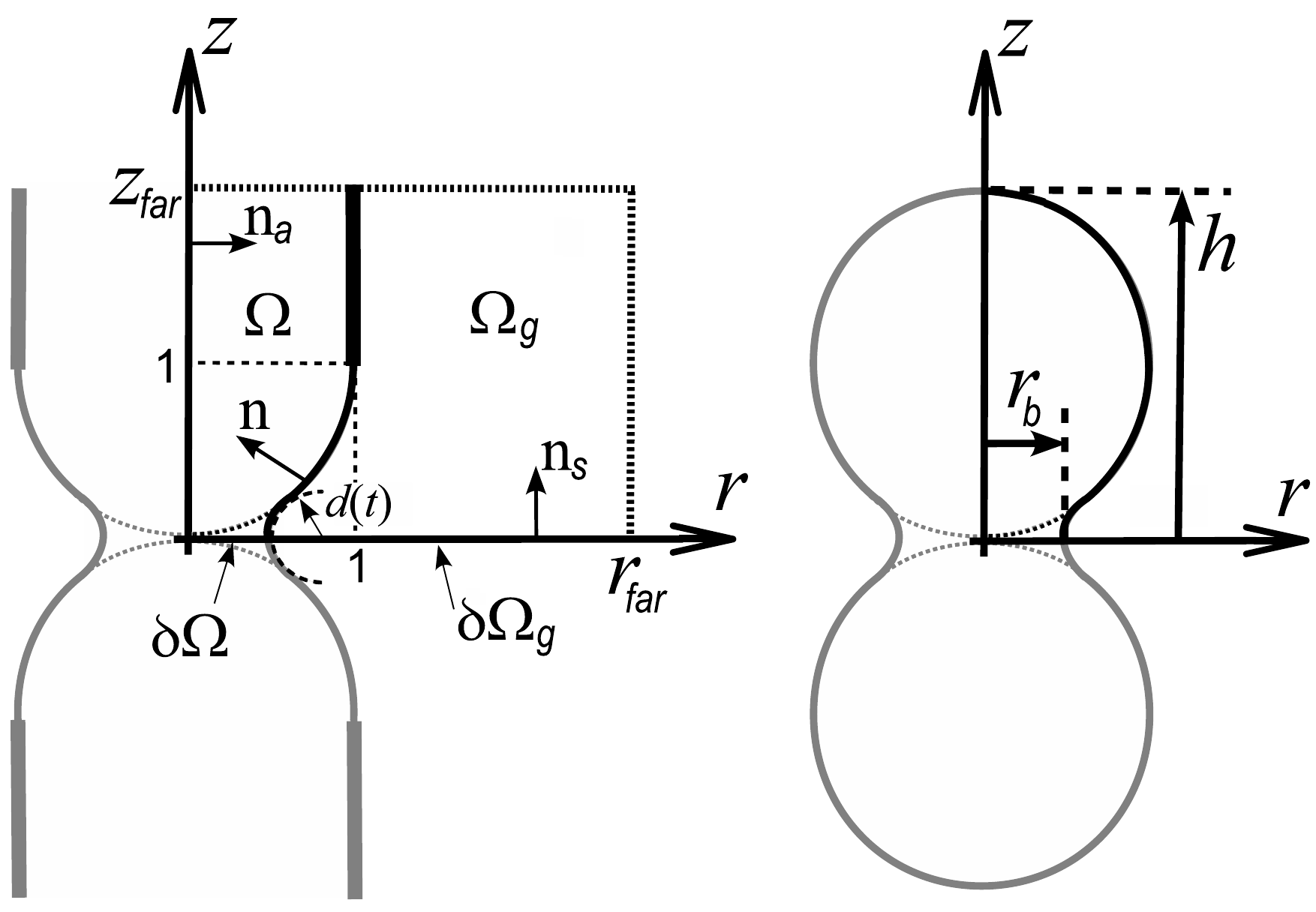}
 \caption{A definition sketch for the coalescence of two identical `pinned hemispheres' grown from syringes (left) and a sketch of coalescing `free spheres' (right) showing the bridge radius $r_b$ and apex height $h$.  In the results section, the bridge radius will simply be denoted as $r$.}
 \label{F:sketch}
\end{figure}

Two different geometries will be considered in this work (Figure~\ref{F:sketch}) both regarding the axisymmetric coalescence of liquid drops formulated in the standard way.  The majority of calculations will be for the typical experimental setup in which hemispherical drops are grown from syringes and surrounded by a viscous gas but at certain points we will also be compelled to study the case of coalescing free spheres.

It has previously been demonstrated that, for the parameter regimes considered, in the initial stages of coalescence the effects
of gravity can be ignored \citep{sprittles_pof2}, so that the
problem becomes symmetric and can be reduced to determining the motion of one drop in the
$(r,z)$-plane of a cylindrical coordinate system with the symmetry conditions on the $z=0$ plane at which the drops initially touch
(Figure~\ref{F:sketch}). The syringe, when considered, is taken to be a
semi-infinite cylinder with zero-thickness walls located at
$r=1,z>1$ which separates the liquid phase $r<1$ from the gas $r>1$,
where the lengths are scaled with the radius of each drop $R$. The precise
far field conditions, i.e.\ those associated with the syringe head,
have a negligible effect on the initial stages of coalescence
\citep{sprittles_pof2}.

Both fluids, i.e.\ the liquid forming the drops and the ambient gas,
are considered to be incompressible and Newtonian with constant
densities $\rho$, $\rho_g$ and viscosities $\mu$, $\mu_g$.  As
before and henceforth, the subscript $g$ refers to properties of the
gas. The fluids occupy domains $\Omega$ and $\Omega_g$, respectively
(Figure~\ref{F:sketch}). To non-dimensionalise the system of the
governing equations for the bulk variables, we use the drop radius
$R$ as the characteristic length scale, $U_{v}$ as the scale for
velocities, $T_{v}$ as the time scale and $\sigma/R$ as the scale
for pressure. Then, the continuity and momentum balance equations
in the two phases take the form
\begin{equation}\label{ns}
\nabla\cdot\mathbf{u} = 0,\qquad Re~\left[\pdiff{\mathbf{u}}{t} + \mathbf{u}\cdot\nabla\mathbf{u}\right] = \nabla\cdot\mathbf{P};\qquad \mathbf{P} = -p\mathbf{I} + \left[\nabla\mathbf{u}+
 \left(\nabla\mathbf{u}\right)^T\right], \qquad \mathbf{r}\in\Omega
\end{equation}
\begin{equation}\label{ns1}
\nabla\cdot\mathbf{u}_g = 0,\qquad \bar{\rho}Re~\left[\pdiff{\mathbf{u}_g}{t} + \mathbf{u}_g\cdot\nabla\mathbf{u}_g\right] = \nabla\cdot\mathbf{P}_g; \qquad \mathbf{P}_g = -p_g\mathbf{I} + \bar{\mu}\left[\nabla\mathbf{u}_g+\left(\nabla\mathbf{u}_g\right)^T\right], \qquad \mathbf{r}\in\Omega_g
\end{equation}
where $\mathbf{P}$, $\mathbf{u}$ and $p$ are the stress
tensor, velocity and pressure in the fluid; $\mathbf{I}$ is
the metric tensor of the coordinate system. The non-dimensional
parameters are the Reynolds number $Re=\rho \sigma R/\mu^2$ based on
the liquid's properties, the gas-to-liquid density ratio
$\bar{\rho}=\rho_g/\rho$ and the corresponding viscosity ratio
$\bar{\mu}=\mu_g/\mu$.

Here, we have assumed that both the liquid and gas are
incompressible so that the Mach number $M=U/a$, where $a$ is the
speed of sound, in each fluid is small throughout the drops' motion.
The fastest speed will be at the bridge front for the coalescence of
the lowest-viscosity drops considered, and the largest Mach number
will be in the air phase, where $a\sim340$ m~s$^{-1}$ as opposed
to the liquid where it is many times larger. A good estimate for
the maximum speed $U$, as confirmed a-posteriori by computations, is
the capillary speed $U_v=\sigma/\mu$ which is a maximum of $20$~m~s$^{-1}$ for the liquids considered giving
in the air phase $M=0.06$. Thus, our assumption of incompressibility
is well justified, especially given that in the well-known
isentropic formulas of gas dynamics the magnitude of the density variation is
proportional to $M^2$.

The conventional boundary conditions used for free-surface flows are
the kinematic condition, stating that the fluid particles forming
the free surface stay on the free surface at all time; the
continuity of both components of velocity across the interface; and
the balance of tangential and normal forces acting on an element of
the free surface from the two bulk phases and from the neighbouring
surface elements:
\begin{equation}\label{ckin}
\pdiff{f}{t} + \mathbf{u}\cdot\nabla f = 0,\qquad
\mathbf{u}_g=\mathbf{u},
\end{equation}
\begin{equation}\label{cstress}
 \mathbf{n}\cdot\left(\mathbf{P}-\mathbf{P}_g\right)\cdot
  \left(\mathbf{I}-\mathbf{n}\mathbf{n}\right) =\mathbf{0},
  \qquad
 \mathbf{n}\cdot\left(\mathbf{P}-\mathbf{P}_g\right)\cdot
 \mathbf{n} =\nabla\cdot\mathbf{n}.
\end{equation}
Here $f(r,z,t)=0$ describes the \emph{a priori} unknown free-surface
shape, with the  unit normal vector $\mathbf{n} = \nabla f/|\nabla
f|$ pointing into the liquid, and the tensor
$(\mathbf{I}-\mathbf{n}\mathbf{n})$ extracts the component of a
vector parallel to the surface with the normal $\mathbf{n}$.

At the plane of symmetry $z=0$, the standard symmetry conditions of
impermeability and zero tangential stress are applied
\begin{eqnarray}\label{csym}
\mathbf{u}\cdot\mathbf{n}_s = 0, \qquad \mathbf{n}_s\cdot\mathbf{P}\cdot \left(\mathbf{I}-\mathbf{n}_s\mathbf{n}_s\right) =\mathbf{0},   \qquad & \mathbf{r}\in\partial\Omega; \\
\mathbf{u}_g\cdot\mathbf{n}_s = 0,\qquad\mathbf{n}_s\cdot\mathbf{P}_g\cdot \left(\mathbf{I}-\mathbf{n}_s\mathbf{n}_s\right) =\mathbf{0}, \qquad & \mathbf{r}\in\partial\Omega_g,
\end{eqnarray}
where $\mathbf{n}_s$ is the unit normal to the plane of symmetry. In
the conventional model we are studying here, the free surface is assumed to always be
smooth so that where it meets the plane of symmetry we have
$\mathbf{n}\cdot\mathbf{n}_s=0$.

On the axis of symmetry $r=0$, the standard normal and tangential
velocity condition state that the velocity has only the component
parallel to the axis and the radial derivative of this component is
zero (the velocity field is smooth at the axis),
\begin{eqnarray}
\mathbf{u}\cdot\mathbf{n}_a = 0,
 \qquad
 \frac{\partial}{\partial r}
 [ \mathbf{u} \cdot (\mathbf{I}-\mathbf{n}_a\mathbf{n}_a) ] =0,
 \qquad & r=0;
\end{eqnarray}
where $\mathbf{n}_a$ is the unit normal to the axis of symmetry in
the $(r,z)$-plane.

For the case of coalescing free spheres, the free surface is assumed smooth at the apex $r=0, z=h(t)$ so that $\mathbf{n}\cdot\mathbf{n}_a=0$ there, whilst the case of coalescing pinned hemispheres requires more conditions to account for the presence of the syringe.  Specifically, at the point in the $(r,z)$-plane where the
(initially hemispherical) free surface meets the syringe tip, we have
a pinned contact-line:
\begin{equation}\label{pinned_shape}
 f(1,1,t)=0 \qquad (t\ge0).
\end{equation}

It is assumed that in the far field, the exterior gas and the liquid
inside the syringe are at rest, so that
\begin{equation}\label{pinned_shape}
\mathbf{u},~\mathbf{u}_g\rightarrow \mathbf{0} \qquad\hbox{as}\qquad r^2+z^2\rightarrow\infty,
\end{equation}
whilst on the cylinder's surface, no-slip is applied
\begin{equation}\label{pinned_shape}
\mathbf{u}=\mathbf{u}_g=0 \qquad\hbox{at}\qquad r=1,z\ge 1.
\end{equation}

The conventional model postulates that, once the drops come into
contact, they produce a smooth free surface, i.e.\ they coalesce on
the sub-fluid-mechanical scale and round the corner enforced by the
drops' configuration at the moment of touching. A bridge of zero
radius with infinite azimuthal and longitudinal curvatures of the
free surface is obviously a singular configuration and hence cannot
be used as a starting point for computation; one has to use an
approximation to this configuration, i.e.\ specify the initial shape
as having, near the origin, a tiny but finite-size bridge with some
radius $r_{min}>0$, where the free surface crosses the plane
of symmetry at a right angle. By introducing explicitly the
radius $r_{min}$ from which our computations start, we ensure that
they are mesh-independent under refinement, unlike those studies in
which the initial bridge radius was defined in terms of the mesh, e.g.\ \cite{menchacarocha01}. Then, we can study the
effect of a finite $r_{min}$ separately.

The free-surface shape far away from the origin (i.e.\ from the
point of the initial contact) is initially the undisturbed
hemispherical/spherical drop. A shape which satisfies these criteria can be
taken from \cite{hopper84}, i.e. the analytic two-dimensional
solution to the problem for Stokes flow. In parametric form, the
initial free-surface shape is taken to be
\begin{eqnarray} \notag
 r(\theta) = \sqrt{2}\left[(1-m^2)(1+m^2)^{-1/2}
 (1+2m\cos\left(2\theta\right)+m^2)^{-1}\right](1+m)\cos\theta, \\  \label{ic2}
 z(\theta) =
 \sqrt{2}\left[(1-m^2)(1+m^2)^{-1/2}(1+2m\cos\left(2\theta\right)+m^2)^{-1}\right]
 (1-m)\sin\theta,
\end{eqnarray}
for $0<\theta<\theta_u$, where $m$ is chosen such that $r(0)=r_{min}$
is the initial bridge radius, which we choose, and $\theta_u$ is
chosen such that $r(\theta_u)=z(\theta_u)=1$ for hemispherical drops and $r(\theta_u)=0$ for spherical ones. Notably, for
$r_{min}\to0$ we have $m\to1$  and $r^2+(z-1)^2=1$, i.e.\ the drop's
profile is a semicircle of unit radius which touches the plane of
symmetry at the origin as required.

An alternative approach, considered briefly in \S\ref{S:ICs}, is to start the simulation with a truncated sphere of radius $r_{min}$ which meets the plane-of-symmetry at an angle $\theta=180^\circ$ and then to rapidly change $\theta$ until a smooth free surface ($\theta=90^\circ$) is obtained.  To do so, one can prescribe the angle $\theta(t) = 180^\circ - 90^\circ \min(1,t/T_r)$ where $T_r$ is the timescale over which the free surface is `rounded'.

Finally, we need to prescribe the fluid initial velocities in the
two phases, which we will assume to be zero:
\begin{equation}\label{ic-u}
 \mathbf{u}= \mathbf{u}_g=\mathbf{0}\qquad\hbox{at }t=0.
\end{equation}
This condition is based on the assumption that the drops are brought together slowly.  Computations confirm that if instead the maximum possible approach velocity $8\times10^{-5}$m~s$^{-1}$ from the experiments in \cite{paulsen11} is used to formulate an initial condition, then the results obtained are graphically indistinguishable from those presented.  This is to be expected as the initial bridge speeds are many times larger than the approach speeds used.

\section{Computational approach}\label{S:comp}

In order to tackle the coalescence phenomenon in its entirety, we
must solve a two-phase free-boundary problem with effects of
viscosity, inertia and capillarity all present, so that a
computational approach is unavoidable.  To do so, we use a
finite-element framework which was originally developed for dynamic
wetting flows and has been thoroughly tested in
\cite{sprittles_ijnmf,sprittles_jcp} as well as being applied to
flows undergoing high free-surface deformation in
\cite{sprittles_pof}, namely microdrop impact onto and
spreading over a solid surface. Notably, the method has been
specifically designed for multiscale flows, so that the very small
length scales associated with the early stages of coalescence can be
captured alongside the global dynamics of the two drops' behaviour.
In other words, all of the spatio-temporal scales which are resolved
in the electrical experiments mentioned earlier \citep{paulsen11},
as well as the scales associated with later stages of the drop's
evolution, which are accessible to optical observation, can, for the
first time, be simultaneously resolved. A user-friendly step-by-step
guide to the implementation of the method has already been provided
\citep{sprittles_ijnmf,sprittles_jcp} and, although this is for a
single-phase flow, the extension to a two-phase flow is a relatively
straightforward procedure which doesn't introduce any conceptually
new ideas to the framework already used.  This code has also been
benchmarked in \cite{sprittles_pof2} against previous simulations of
coalescence in \cite{paulsen12} at the scales resolved in that work.

The computational domain is truncated, so that `far-field'
conditions on the gas and the liquid in the cylinder must be applied
at a finite distance from the origin.  To do so, we apply `soft'
conditions on these boundaries and ensure that these boundaries are
sufficiently far from the coalescing hemispheres that neither the
conditions specified there nor any further increase of $r_{far}$ and
$z_{far}$ (Figure~\ref{F:sketch}) have any influence on the drops'
dynamics.

\section{Parametric Study}\label{S:para}

A systematic study of the governing parameters in the coalescence process will now be considered and then, in \S\ref{S:exp}, the results will be compared to available experimental data. An advantage of this approach is that the parameters can be independently varied in the computations whereas in the experiments often it is the viscosity which is varied, so that $Re$ and $\bar{\mu}$ are related, which makes isolating the effect of each parameter more difficult.  Our approach here will be to consider the simplest possible setup first, and then add layers of complexity.  For example, first of all a passive gas will be considered ($\bar{\mu}=\bar{\rho}=0$), and only once the role of the remaining parameters has been established will the gas dynamics be considered.  Once the full parametric study and comparison to experiment have been performed, this will all be tied together with the published literature in \S\ref{S:disc}.

To understand the different regimes of drop coalescence, the appropriate scalings in these regimes, the crossover between them and their comparison to experiments which are able to capture many decades of bridge radius, results will be given on log-log plots.

\subsection{Influence of initial conditions, dimensionality and geometry}\label{S:ICs}

To estimate the influence of the initial conditions compared to the solution obtained as the initial bridge radius $r_{min}\rightarrow0$, computations for finite $r_{min}$ are compared to Hopper's solution (\ref{hopper}) which was obtained for the inertialess coalescence of two-dimensional liquid cylinders.  The possible effect of errors associated with a finite initial radius is particularly important when considering the initial stages of motion where small changes in the initial time can sometimes drastically alter the agreement between experiments and scalings, see \cite[\S5.3]{thoroddsen05}.  To be consistent with Hopper's solution, we will take $Re=0$ and consider the gas to be passive.

\subsubsection{Effect of finite minimum radius}

\begin{figure}
     \centering
\includegraphics[scale=0.3]{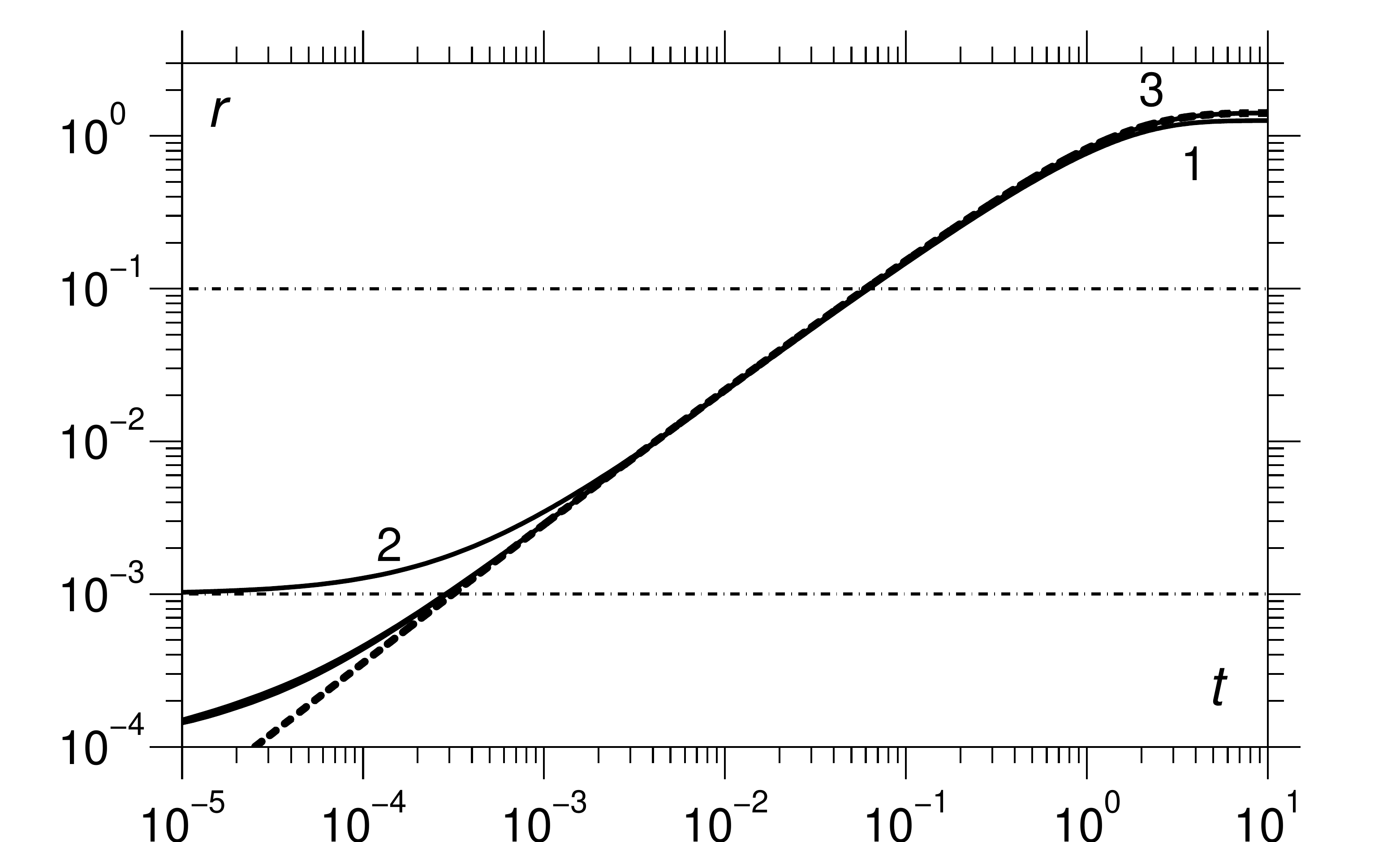}
 \caption{Bridge radius as a function of time for the case $Re=0$ compared to Hopper's solution (\ref{hopper}), the dashed line. Curve 1 is for three-dimensional free spheres with $r_{min}=10^{-4}$, curve 2 the same except that $r_{min}=10^{-3}$ and curve 3 is for two-dimensional free cylinders ($r_{min}=10^{-4}$).}
 \label{F:2d_3d}
\end{figure}

Simulations shown in Figure~\ref{F:2d_3d}, performed for $r_{min}=10^{-4}$, show that the computed solution (curve 3) for the bridge radius of free cylinders coalescing is graphically indistinguishable from Hopper's exact solution (dashed line) from $r=10^{-3}$ (marked by the lower horizontal dash-dot line) onwards.  This is despite the fact that in Hopper's solution at $t=0$ the bridge radius is infinitesimal whereas in the computations $r=10^{-4}$.  As a consequence of the observed agreement from $r=10^{-3}$, we do not have to concern ourselves with calculating the time $t_0$ at which the bridge would reach a radius $r_{min}$, and then subtract this from the time elapsed in the computation $t$, i.e.\ to plot $r$ against $t-t_0$; instead, we can simply plot computations from $r=10^{-3}$ knowing that the error associated with starting at a finite bridge radius is negligible.

\subsubsection{Equivalence of two-dimensional and three-dimensional solutions}

Although Hopper's solution is strictly valid only for two-dimensional motion, results in \cite{eggers99} and \cite{paulsen12} suggest that this expression may also approximate the initial stages of the axisymmetric three-dimensional solution as well. The curves in Figure~\ref{F:2d_3d} confirm that this is the case: curves 1 and 3 obtained for coalescing spheres and cylinders, respectively, are graphically indistinguishable up to at least $r = 10^{-1}$ (upper horizontal dash-dot line).  Clearly, at longer times the two curves must diverge as the two configurations have different equilibrium bridge radii $r_{eqm}$, with $r_{eqm}=2^{1/2}=1.41$ for cylinders and $r_{eqm}=2^{1/3}=1.26$ for spheres.

To re-enforce our arguments about the effect of the initial bridge radius, computations for free spheres with a larger $r_{min}=10^{-3}$ are shown by curve 2 in Figure~\ref{F:2d_3d} and it can be seen that in this case after $r=10^{-2}$ the curve falls on top of the computed solution for $r_{min}=10^{-4}$ (curve 1) and hence also Hopper's solution.  Thus, for both $r_{min}$ considered, at $r=10r_{min}$ the curves are insensitive to the finite initial radius used.  Notably, computations confirm that for the range of $Re$ considered in this work, similar levels of insensitivity to the initial finite radius were observed.

\subsubsection{Effect of geometry}

\begin{figure}
     \centering
\includegraphics[scale=0.3]{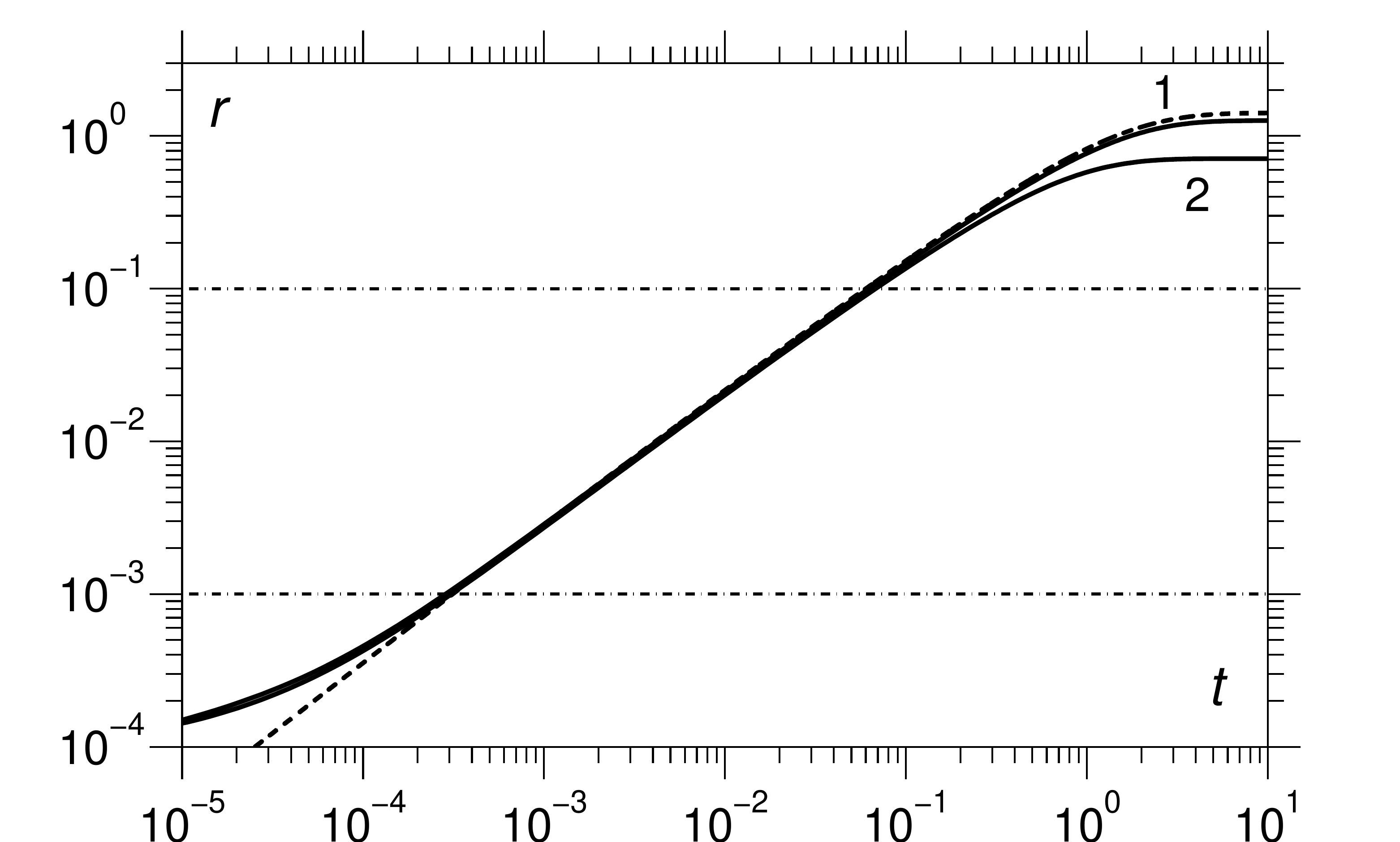}
 \caption{Bridge radius as a function of time for the case $Re=0$ for free spheres for (curve 1) and pinned hemispheres (curve 2). The dashed lined is Hopper's solution (\ref{hopper}).}
 \label{F:geometry}
\end{figure}

In Figure~\ref{F:geometry}, the evolution of the bridge radius for the coalescence of three-dimensional free spheres (curve 1) and pinned hemispheres (curve 2) is shown.  Very slight deviations between the two curves are observed for the entire time; however, until $r =10^{-1}$ these differences are so small that they are likely to fall below the resolution of any experimental accuracy.  Therefore, the effect of geometry can be considered negligible until $r = 10^{-1}$ after which the bridge of pinned hemispheres is slower as it asymptotes to a smaller equilibrium radius of $r_{eqm}=0.71$ than the free spheres ($r_{eqm}=1.26$).

Notably, the case $Re=0$ is most likely to highlight any effect of the global geometry (far away from the bridge) on the initial stages of the bridge's evolution as at finite Reynolds number, as we will see later, the flow near the bridge will be more `localised' in comparison to Stokes flow, where the entire body of fluid moves from $t=0$.

\subsubsection{Effect of initial free surface shape}

To further re-enforce the point, that the effect of our initial conditions is negligible from $r=10^{-3}$, we have compared the two different start-up strategies proposed in \S\ref{S:model}, namely to either (a) use Hopper's solution as an initial condition for the free surface shape or (b) use a truncated sphere and make the free surface smooth where it meets the plane of symmetry over a time-scale $T_r$, which we choose here to be $T_r = 10^{-5}$.  Again, from $r=10^{-3}$, the curves obtained from either start-up strategy were seen to be graphically indistinguishable.

\subsubsection{Summary}

For $r_{min}=10^{-4}$, from $10^{-3}<r<10^{-1}$, i.e.\ what will be considered as the `initial stages of motion', the bridge evolution of the coalescing drops is graphically indistinguishable:
\begin{itemize}
  \item From those obtained for $r_{min}=0$.
  \item For spheres and cylinders of the same radius.
  \item For free spheres and pinned hemispheres
\end{itemize}

\subsection{Effect of the Reynolds number}\label{S:Re}

If the parameters governing the initial configuration are fixed, and the gas is still passive, then the only parameter remaining is the Reynolds number $Re$. Unless specified, computations are with pinned hemispheres, which in all cases considered give the same behaviour as free spheres up to at least $r=0.1$.

\subsubsection{Small Reynolds numbers: $Re \leq 1$}

\begin{figure}
     \centering
\includegraphics[scale=0.3]{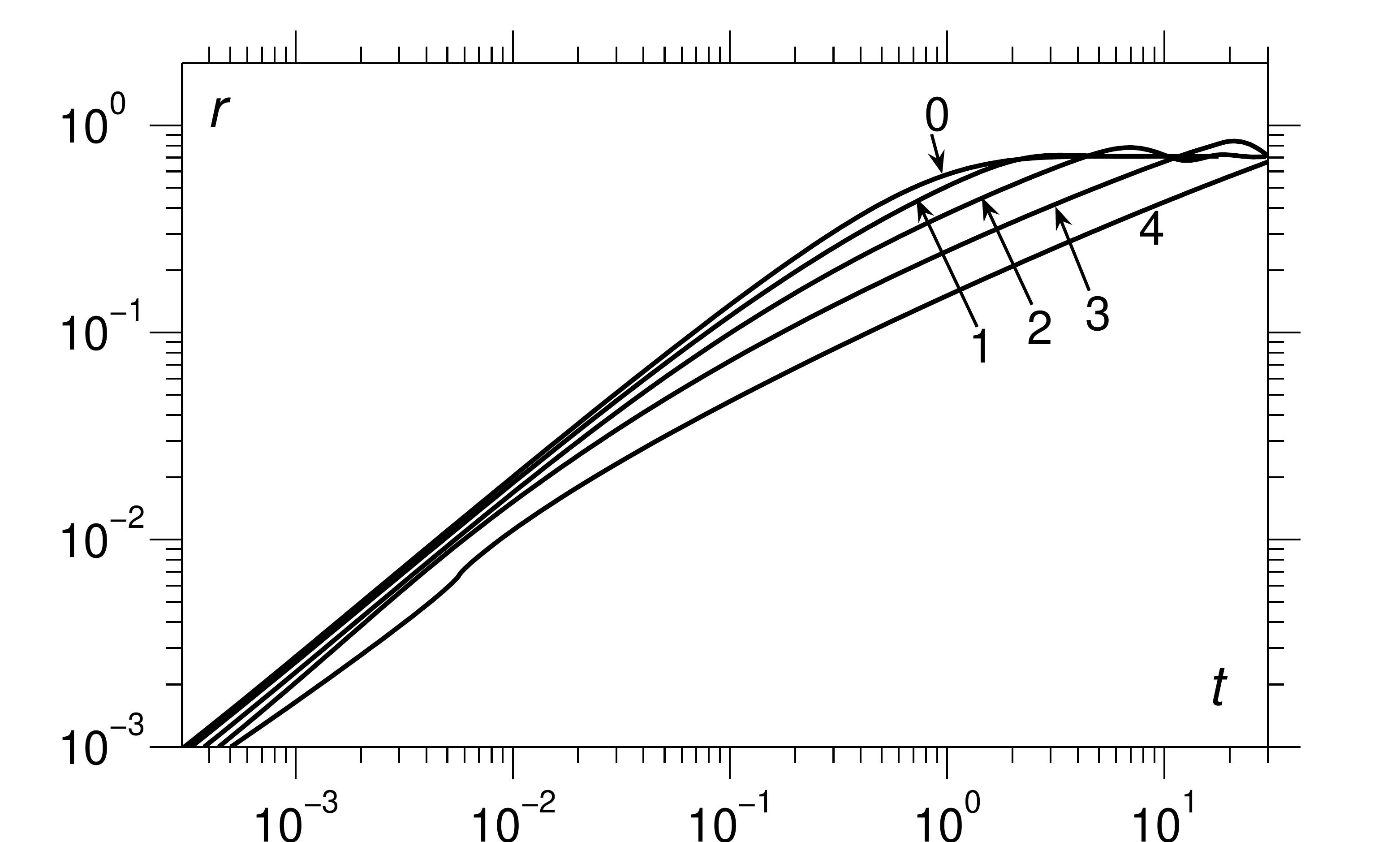}
 \caption{Bridge radius as a function of time for a variety of different Reynolds numbers. Curve 0 is for $Re=0$ (curves for $Re\leq1$ are graphically indistinguishable from it), 1: $Re=10^1$, 2: $Re=10^2$, 3: $Re=10^3$ and 4: $Re=10^4$. }
 \label{F:Re}
\end{figure}

\emph{All} curves for $Re \leq 1$ are seen to be graphically indistinguishable on a log-log plot from those obtained for $Re=0$ in Figure~\ref{F:Re}. This is an intriguing result: measurements of the bridge radius show no evidence of an ILV regime for $Re\leq1$. 

\begin{figure}
     \centering
\includegraphics[scale=0.3]{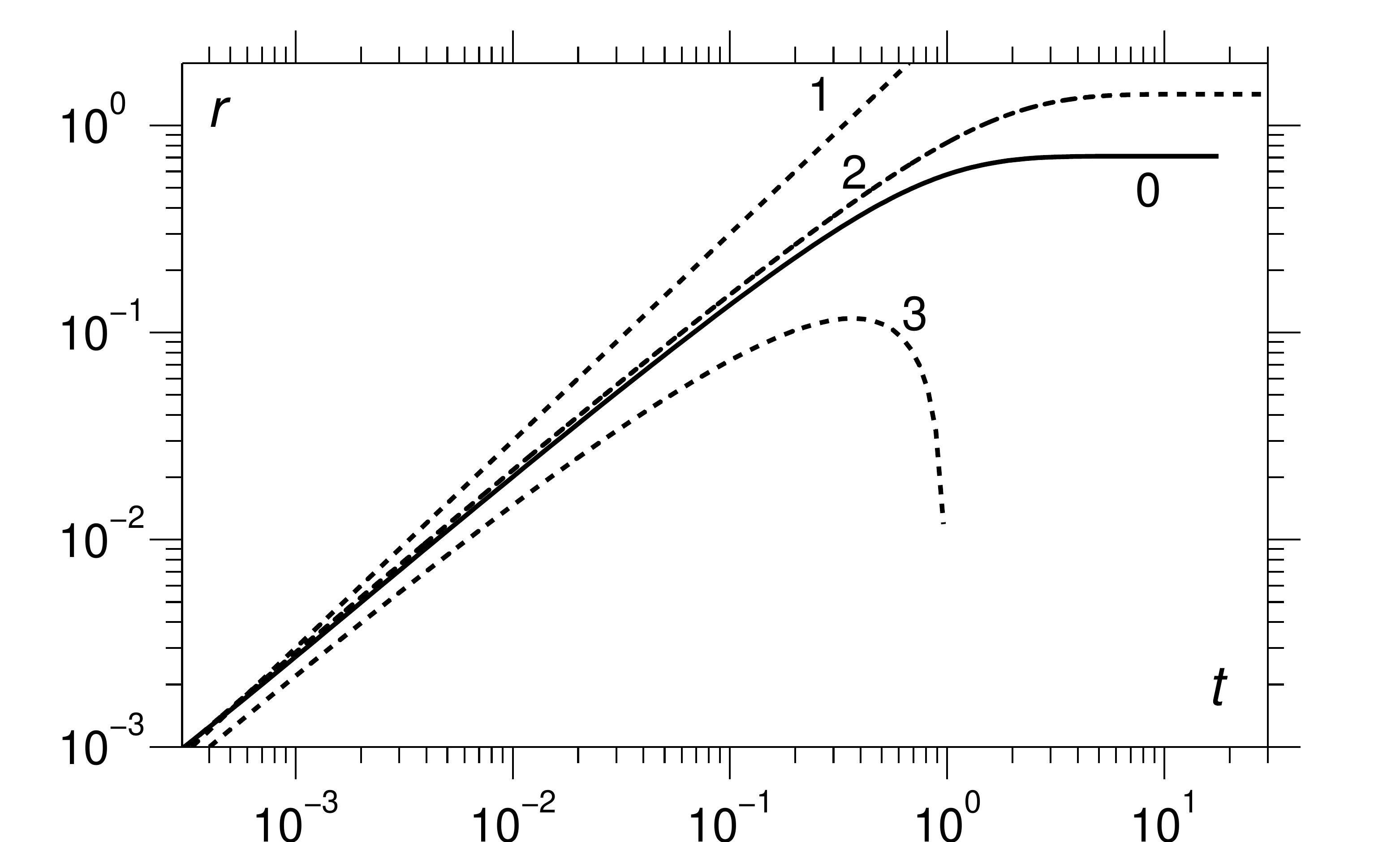}
 \caption{Comparison of the full numerical solution for $Re=0$ (and hence all $Re\leq1$) curve 0 with scalings and the exact solution for free cylinders.  Curve 1: linear plot (\ref{ilv}) of $r=3t$; curve 2: Hopper's solution for free cylinders (\ref{hopper}); curve 3: formula (\ref{eggers}) for a passive gas, i.e.\ $r=-(1/\pi)t\ln t$.  It is clear that Hopper's solution best approximates the full numerical solution.}
 \label{F:scalings}
\end{figure}

As can be seen from curve 2 in Figure~\ref{F:scalings}, for $Re\leq 1$ it is Hopper's exact solution (\ref{hopper}) that provides the best approximation of the computed bridge front evolution for $r<0.1$, confirming again that this range is described by inertialess Stokes flow.  In other words, we are in what has classically been referred to as a `viscous regime'. Curve 3 is the expression (\ref{eggers}) from \cite{eggers99}, which is an asymptotic approximation of Hopper's solution (curve 2).  It is seen to be inaccurate in the range $10^{-3}<r<10^{-1}$ of interest, as suggested in \cite{eggers99} where $r<0.03$ is said to be the range of applicability of their formula. In a previous work \citep{sprittles_pof2}, this expression was shown to describe reasonably the conventional model when $C_v = C_v(Re)$ in (\ref{eggers}) was \emph{fitted}, which, strictly speaking, it should not be, as (\ref{eggers}) was originally derived as an approximation to the exact expression (\ref{hopper}).  Notably, the linear expression (curve 1), indicative of the ILV regime, is also seen to diverge from the computed result (curve 0).

\subsubsection{Large Reynolds numbers: $Re \geq 1$}

From Figure~\ref{F:Re}, it can be seen that curve 1, for $Re=10$, has diverged noticeably from the Stokes flow solution (curve 0) by around $r=0.1$.  A further increase in the Reynolds number to $Re=100$ (curve 2) ensures no agreement with the Stokes solution, although the divergence is rather small for $r<0.01$.  Clearly, once $Re\geq 10^3$ significant deviations from the Stokes flow solution are seen, so that inertial effects are becoming increasingly important.

\begin{figure}
     \centering
\includegraphics[scale=0.3]{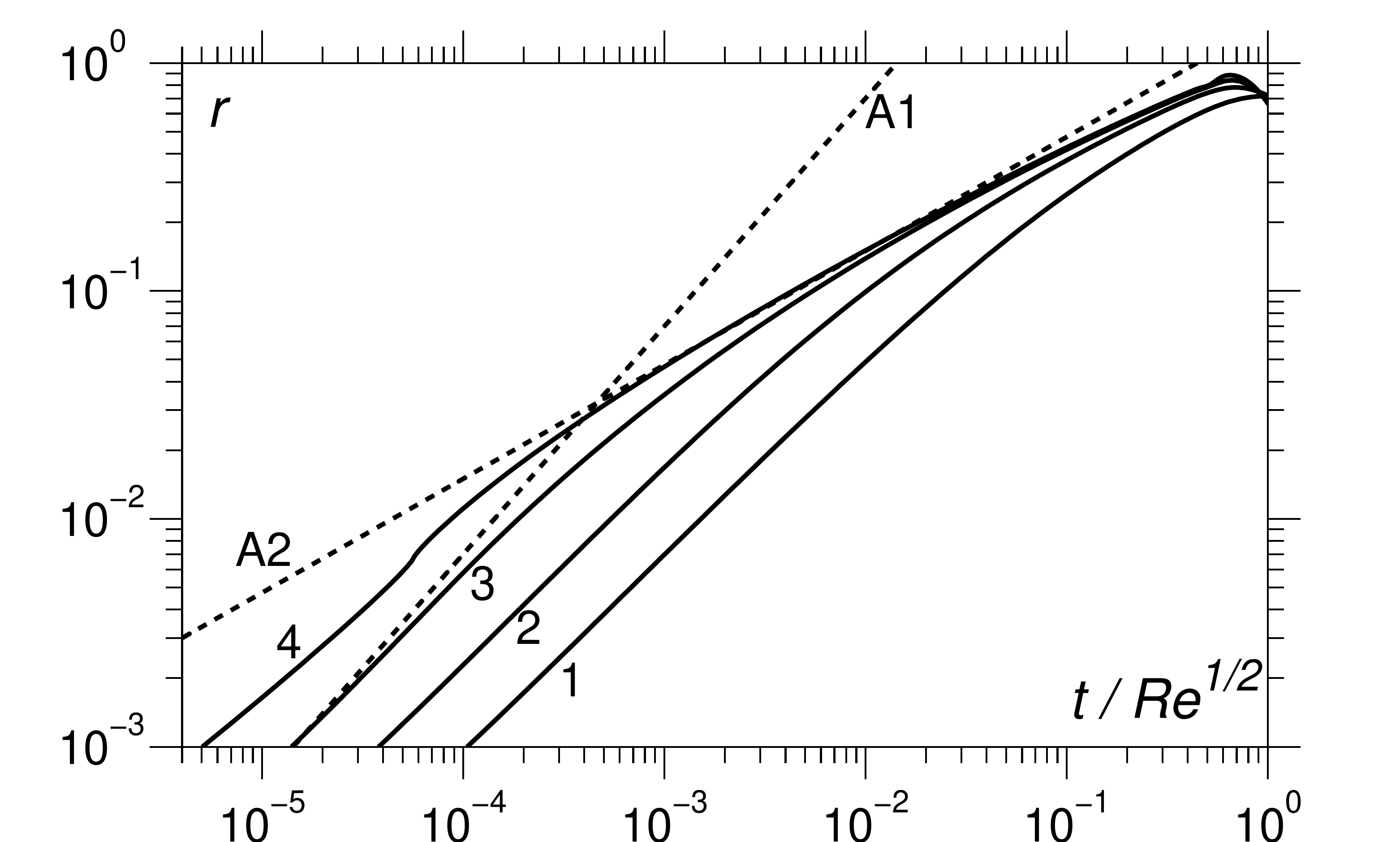}
 \caption{Curves from Figure~\ref{F:Re} plotted against the time scaled using $T_i$, i.e.\ $r$ against $t_i=t/Re^{1/2}$, characteristic of the inertia-dominated regime as opposed to $T_v$ in Figure~\ref{F:Re}. Curve 1: $Re=10^1$, 2: $Re=10^2$, 3: $Re=10^3$ and 4: $Re=10^4$. The dashed line $A1$ is for (\ref{ilv}), i.e. a linear curve ($r=70t_i$) and $A2$ is for (\ref{io}), i.e. the scaling $r=1.5 t_i^{1/2} = 1.5t^{1/2}/Re^{1/4}$}
 \label{F:Re_inertial}
\end{figure}

Given that the inertial regime is characterised by a different time scale $T_i=(\rho R^3/\sigma)^{1/2}$ as opposed to $T_v=\mu R/\sigma$, in Figure~\ref{F:Re_inertial} we plot the curves of Figure~\ref{F:Re} against the (dimensionless) inertial time $t_i = t (T_v/T_i) = t/Re^{1/2}$ instead of the viscous one.

The `inertial regime' itself is usually characterised by the scaling in (\ref{io}), and by fitting the prefactor ($C_i = 1.5$) to the curve from the highest Reynolds number considered (curve 4), we obtain the dashed line $A2$ in Figure~\ref{F:Re_inertial}. One can see that at $Re=10^4$ (curve 4), the inertial scaling (curve $A2$) approaches the full numerical solution at around $r = 10^{-2}$, which is consistent with the inertial regime being entered when $r\sim Re^{-1/2}$. For the case of $Re=10^3$ (curve 3), fitting (\ref{ilv}) to the early time behaviour gives dashed line $A1$, so that if the crossover is defined where curve $A2$ meets curve $A1$, as considered in \cite{paulsen11}, this will occur at around $r\sim2\times10^{-2}$, i.e.\ again at $r\sim Re^{-1/2}=10^{-3/2}=3\times10^{-2}$. The details of this crossover will be considered in far greater details in \S\ref{S:master}.

Notably, the scaling (\ref{io}) has a rather limited region of applicability, even when $Re$ is sufficiently large to ensure the drops are in an `inertial regime'. This aspect is considered in detail in \cite{sprittles14_pre}, where an improved scaling law for this regime is derived and shown to agree well with both the fully-computed solution as well as a range of experimental data from the published literature.

For $Re=10^2$ (curve 2) and $Re=10^1$ (curve 1), both (\ref{hopper}) and (\ref{io}) fail to approximate any of the observed behaviour, meaning that this is a region of parameter space where both the viscous and inertial forces are important, so that simplified expressions based on neglecting either of these will be inherently inaccurate.  This area of parameter space will later be referred to as the `transition' region and will be rigorously defined in \S\ref{S:disc}.


\subsubsection{Global motion and the ILV regime}

It has been shown that for $Re\leq 1$, the early stage of the bridge's propagation is well approximated by Hopper's solution (\ref{hopper}) for free cylinders whilst for $Re>1$ all of the proposed expressions fail to describe the initial stages of motion.  At first sight, this appears to contradict the results of \cite{paulsen12}, where it was shown that for $Re\leq 1$ ($Oh>1$), an ILV regime is present where inertia cannot be neglected, so that (\ref{hopper}) does not hold.  Simulations and experiments on free spheres confirmed, by measuring the speed at which the centres of the drops move towards each other, that the Stokes flow solution does not accurately describe the global motion of the drops in the very initial stages of motion.

\begin{figure}
     \centering
\includegraphics[scale=0.3]{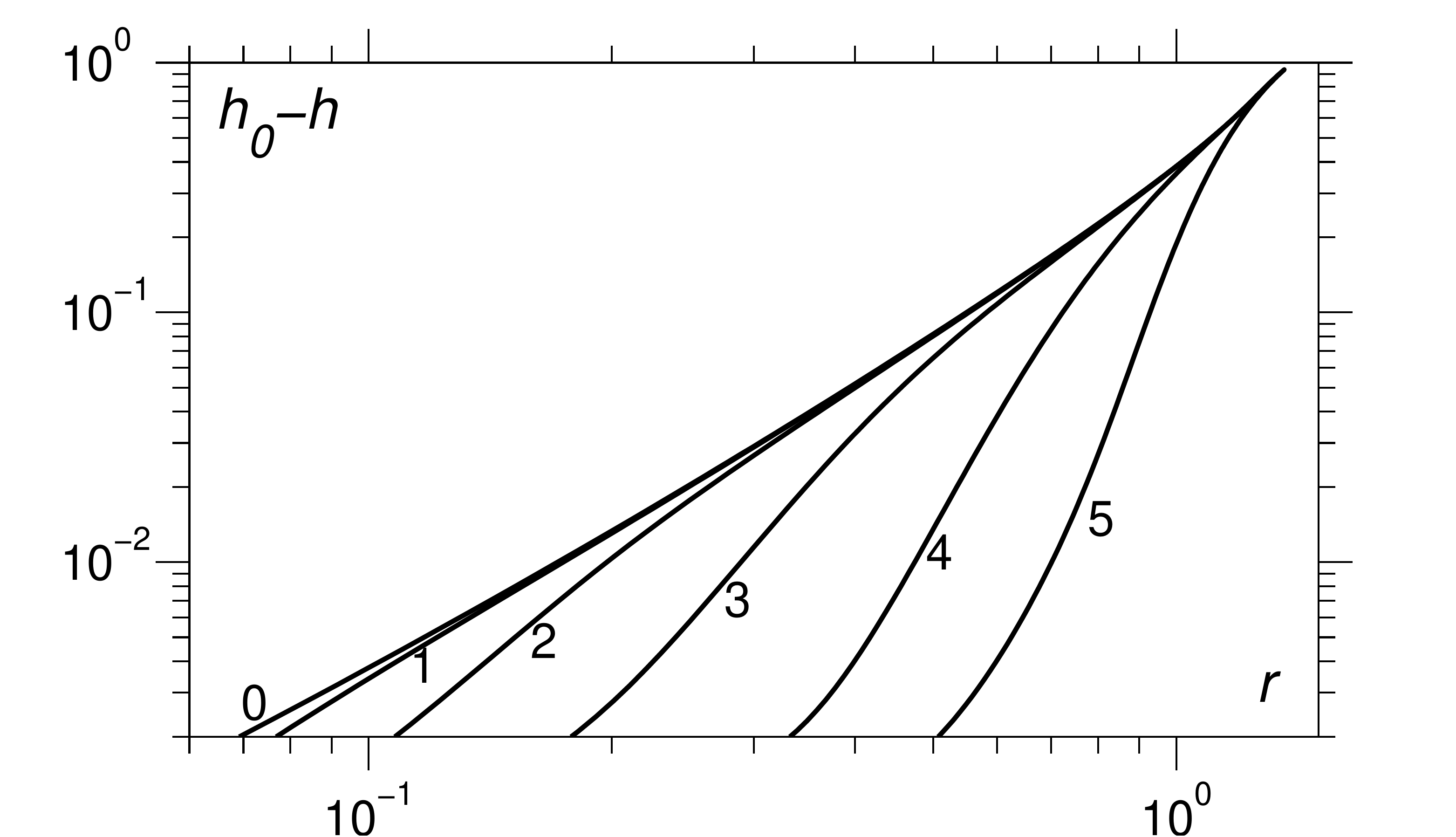}
 \caption{Distance $h_{0}-h$ that the apex has moved plotted against bridge radius $r$ for different Reynolds numbers. Curve 0: $Re=0$, curve 1: $Re=10^{-3}$, 2: $Re=10^{-2}$, 3: $Re=10^{-1}$, 4: $Re=1$, curve 4: $Re=10$ and curve 5: $Re=10^2$.}
 \label{F:height}
\end{figure}

In Figure~\ref{F:height}, we plot the results of simulations performed using free spheres, showing  the distance which the apex height $h$ of the drop (Figure~\ref{F:sketch}) has moved from its initial position $h_0=2$ as a function of bridge's radius.  This is a measure indicative of the influence of the coalescence dynamics on the three-dimensional \emph{global} motion and thus Hopper's two-dimensional solution cannot provide an approximate expression for the evolution of $h$.  From Figure~\ref{F:height}, we can see that even at very small Reynolds number, the curves do not immediately fall onto the computed Stokes flow solution (curve 0), as could have been anticipated from the results in \cite{paulsen12}.  For example, at $Re=10^{-3}$ (curve 1), the bridge travels as far as $r=0.1$ before it approaches the Stokes flow solution. For $Re=10^{-2}$ (curve 2), it is $r=0.2$ and for $Re=10^{-1}$ (curve 3) it takes until $r=0.6$.  At higher Reynolds number, the Stokes flow solution is not approached until the drop starts to reach its equilibrium state.

Thus, as first observed in \cite{paulsen12}, for $0< Re\leq1$ it takes a certain time until the global motion of the drops is approximately described by the Stokes flow solution.  However, during this period, the bridge radius is described perfectly by the Stokes flow solution.  Therefore, we find that the ILV regime is a description of the \emph{global} motion of the drops, as opposed to being an expression for the \emph{local} bridge front evolution, as originally suggested. This means that there is a boundary layer around the bridge front region, which grows in time and inside which the flow is inertialess. Outside this region it is the inertial effects that are important, so that the bridge evolution can be described by Stokes flow solution whilst the global motion of the drops takes some time to follow this behaviour.
%
%

\subsubsection{Summary}

It has been shown that in the initial stages of coalescence ($10^{-3}<r<10^{-1}$) described in the framework of the conventional model, for $Re\leq 1$:
\begin{itemize}
  \item The bridge propagation is described by (\ref{hopper}), i.e.\ by Stokes flow theory.  Neither (\ref{eggers}) nor (\ref{io}) are accurate.
  \item The ILV regime describes the global motion and can be observed by monitoring the motion of the apex of free spheres.
\end{itemize}

For $Re>1$:
\begin{itemize}
 \item For $Re\leq 10^2$ a truly inertial regime, with $C_i$ in (\ref{io}) fixed, is never reached.
  \item For $Re\geq 10^3$ an inertial regime is entered when $r\sim Re^{-1/2}$ after which $r\simeq 1.5t_i^{1/2}$.  
\end{itemize}

\subsection{The influence of a viscous gas}

Having established the role of the parameters for coalescence in a passive gas, we now consider how a dynamically-active viscous gas will effect the process.  To estimate reasonable parameter values, consider typical liquids with viscosities $\mu\sim10^{-3}$--$10$~Pa~s and densities $\rho\sim10^3$~kg~m${}^{-3}$, in contact with gases at atmospheric pressure having $\mu_g\sim10$~$\mu$Pa~s and $\rho_g\sim 1$~kg~m${}^{-3}$.  Then $\bar{\mu}\sim 10^{-6}$--$10^{-2}$ and $\bar{\rho}\sim 10^{-3}$.

\subsubsection{Toroidal bubbles: suppression of their formation by a viscous gas}

\begin{figure}
     \centering
\includegraphics[scale=0.3]{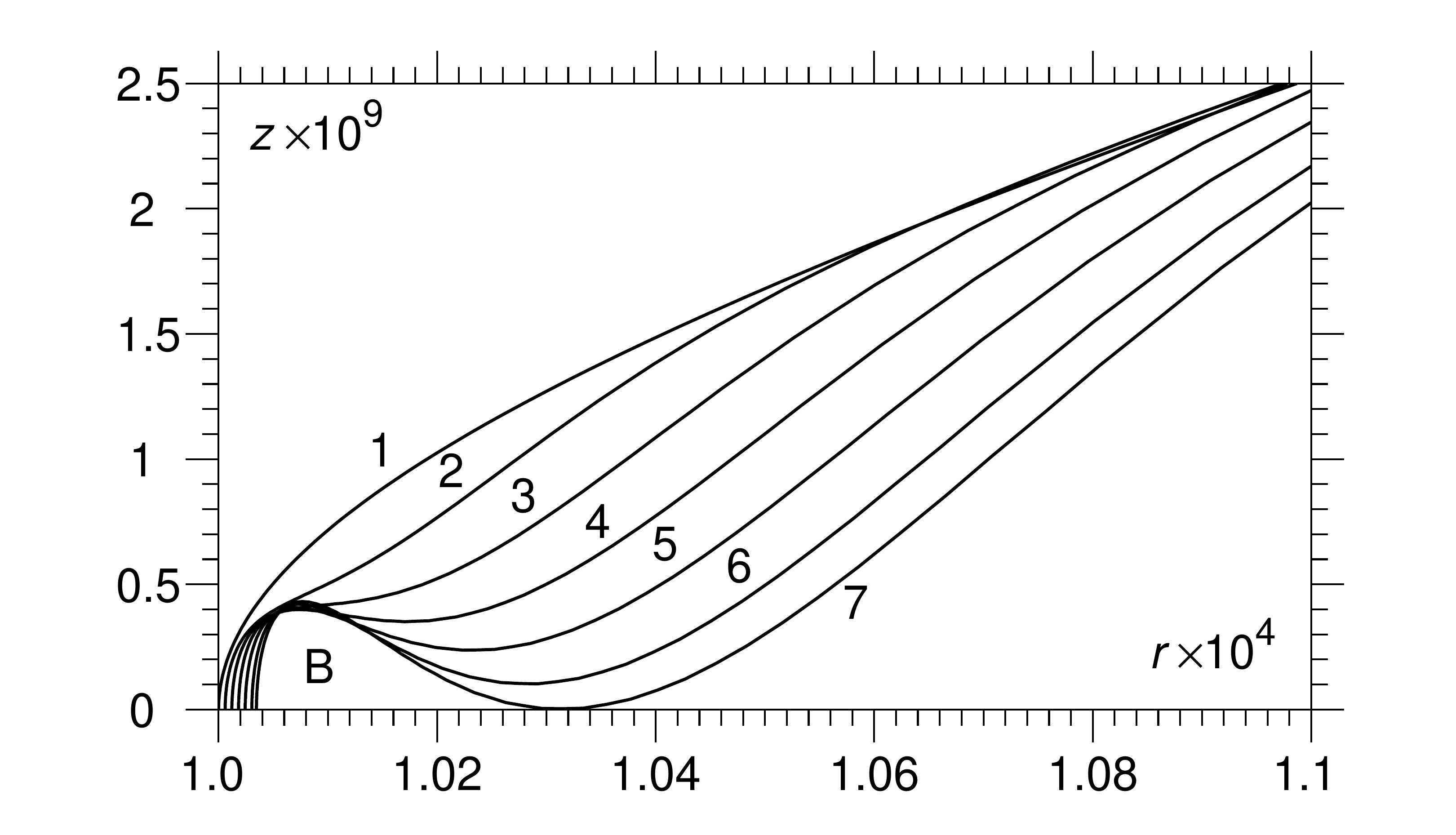}
 \caption{Free surface profiles for the case of two low-viscosity
 liquid drops ($Re=10^4$) coalescing in a \emph{passive} gas, calculated in the framework of the
conventional model, showing the formation of a toroidal bubble (recalling that $z=0$ is a plane of symmetry) at location `B'. Curves 1-6 are in equal time steps from $t=0$ through to $t=10^{-7}$ with curve 7, at the time when the free surface touches the plane of symmetry ($z=0$), at $t=1.14\times10^{-7}$.  This
bubble does not form if the dynamics of the viscous ambient gas is
accounted for.}
 \label{F:toroidal_bubs}
\end{figure}

Before examining the quantitative effect which a viscous gas has on
the propagation of the bridge front, we will look at the qualitative
behaviour of the system in the early stages of coalescence of
low-viscosity drops, where toroidal bubbles have been obtained in
local inviscid boundary-integral calculations
\citep{oguz89,duchemin03}.  A trail of toroidal bubbles are formed at high $Re$ when capillary
waves generated by the disturbance to the free-surface shape caused
by the bridge propagation have a large enough amplitude to reconnect
in front of the bridge (Figure~\ref{F:toroidal_bubs}).  Notably, although the bubble formed in Figure~\ref{F:toroidal_bubs} for $Re=10^{4}$, located at `B', has microscopic dimensions for typical drop sizes, this bubble, should it appear, is likely to be the first in a trail of bubbles of increasing size, as shown in \cite{duchemin03}, so that the question as to whether or not this initial bubble forms is indicative of whether or not macroscopic bubbles could be generated and experimentally detected. In fact, the end of the toroidal bubble formation stage is indicated by a slight `kink' in curve 4 of Figure~\ref{F:Re} (at $r,t\sim10^{-2}$), which disappears when the gas' viscosity is accounted for (c.f.\ curve 3 in Figure~\ref{F:mubar1}). As explained in \cite{sprittles_pof2}, current computational approaches do not accurately capture toroidal bubble formation, but since these bubbles, as shown below, are unphysical, there is little motivation to develop the advanced techniques required to do so.

As the predicted toroidal bubbles have never been observed experimentally, it is of
particular interest to see if the presence of a viscous gas is able
to suppress their formation. This cannot be inferred from previous works which consider either no inertial effects, so that there is no mechanism for bubble formation \citep{eggers99}; no viscous effects, so that bubble formation cannot be suppressed by the gas \citep{duchemin03}; or no gas dynamics at all, as in previous computational works \citep{sprittles_pof2,paulsen12}.

\begin{figure}
     \centering
\includegraphics[scale=0.3]{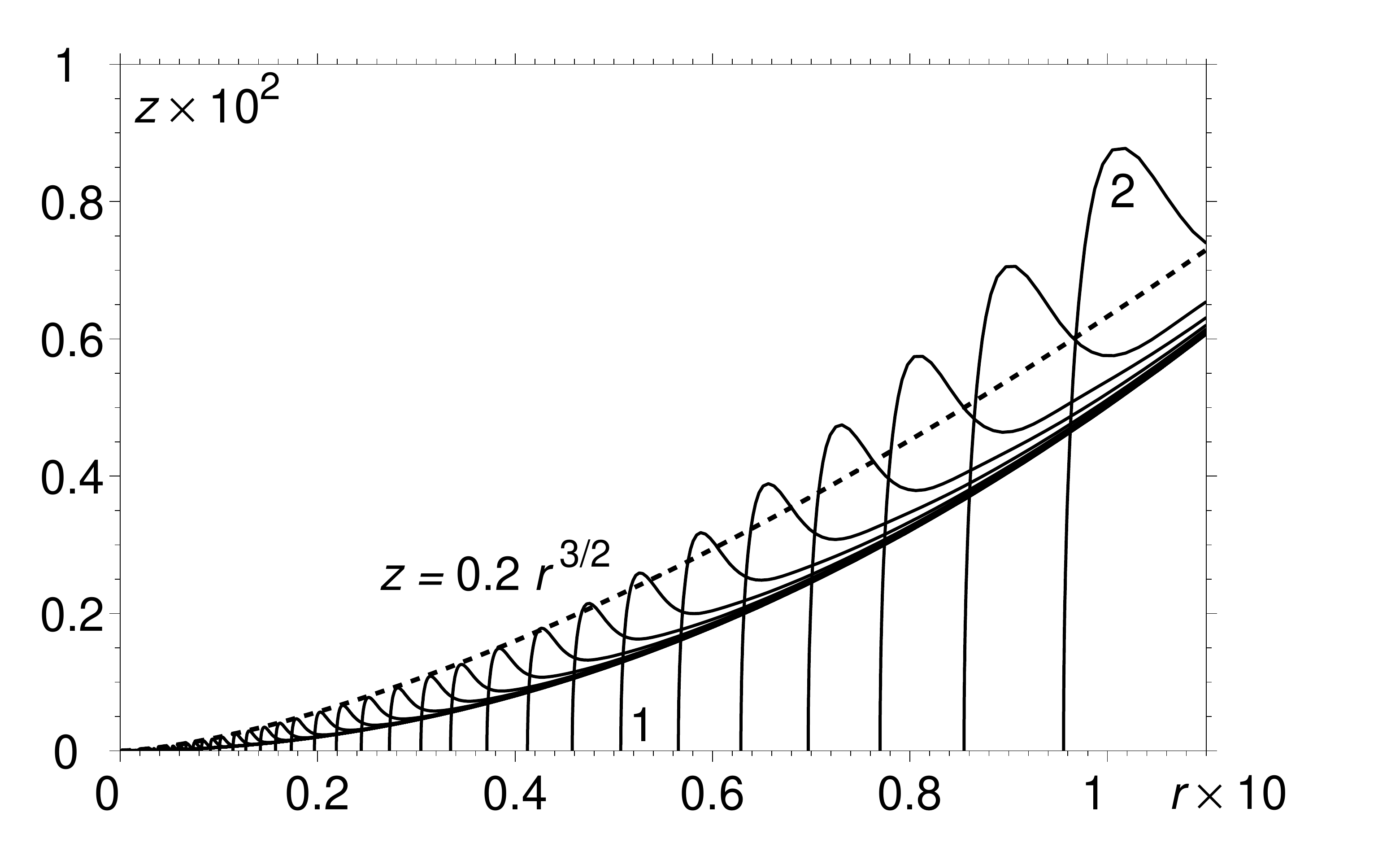}
 \caption{Free surface profiles for the case of two low-viscosity
 liquid drops ($Re=10^4$) coalescing in a viscous gas with
$\bar{\mu}=6\times10^{-3}$ calculated in the framework of the
conventional model. The dashed line shows the scaling for the height
(i.e.\ distance from the plane of symmetry) of the gas bubble
predicted in \cite{eggers99} as a function of bridge front radius.
Note that the scales in the $r$- and the $z$-directions are
different, so that the cross-section of the  trapped air bubble in
front of the bridge is actually more circular than it appears here.
The profiles correspond to different times from the onset of
coalescence; curve~1 correspond to $t = 2.3\times 10^{-2}$ and
curve~2 to $t = 3.5\times 10^{-2}$, with equal time-spacing in
between and outside.}
 \label{F:toroidal}
\end{figure}

In Figure~\ref{F:toroidal}, we show the results of calculations for
the coalescence of low viscosity drops
(also $Re=10^{4}$) in air ($\bar{\mu}=6\times10^{-3}$). As
one can clearly see, a viscous gas acts as a barrier to toroidal
bubble formation, which results in an entirely different behaviour
of the free surface from that previously observed for a passive gas, i.e.\ physically a vacuum, where toroidal
bubbles are formed (Figure~\ref{F:toroidal_bubs}). It can be seen
that the propagating bridge creates a capillary wave and pushes a
gradually growing pocket of air in front of itself, and it is the
dynamics of this pocket of air that now prevents the free surface of
each of the drops from reaching the plane of symmetry, reconnecting,
and trapping a toroidal bubble of air.

As can be seen, the computed free-surface shape is consistent with
the predictions in \cite{eggers99} that  the radius of the curvature
at the bridge front scales like $r^{3/2}$, in contrast to the case
of coalescence in a passive gas,
where the radius of curvature scales like $r^3$. Notably, for the
case of a viscous gas, the radius of curvature at the bridge front
is larger than the undisturbed free-surface height, which scales
like $r^2$, so that the gas bubble protrudes `into' the liquid drop
and causes a local maximum in the free-surface height $z=z(r)$. As
shown in Figure~\ref{F:toroidal} by the dashed line, the latter
scales as $r^{3/2}$ for a considerable distance.

Notably, for realistic parameters it is the viscosity of the ambient
gas that plays the key role in the suppression of the toroidal
bubble appearance. This is highlighted by the fact that, if we set
$\bar{\rho}=0$, toroidal bubbles are not formed until the
gas-to-liquid viscosity ratio is reduced to
$\bar{\mu}\approx10^{-7}$.  Therefore, in reality it is always the viscosity
and not the density of the ambient gas that holds the key to the
toroidal bubble suppression. Indeed, under normal conditions, the
viscosity of the gas is above a certain value, say, 1~$\mu$Pa~s, so
that for the gas-to-liquid viscosity ratio to be of the order of
$10^{-7}$, one must have a liquid with viscosity of the order of
10~Pa~s, and, as shown in \cite{sprittles_pof2}, even for coalescing drops of much lower viscosity than 10~Pa~s, the toroidal bubble does not form
even if the ambient fluid is a vacuum. The same point can be made in
another way: if we take two drops of a low-viscosity liquid that
would produce a toroidal bubble in a vacuum and replace the vacuum
with a gas of gradually increasing viscosity and density, the gas'
viscosity would prevent the bubble formation long before the
gas-to-liquid density ratio has a noticeable effect on the process.

Having established that the presence of a viscous gas completely
alters the initial stages of the coalescence process for a
low-viscosity liquid, it is of interest to study how the parameters
associated with the gas, namely the density and viscosity ratios,
affect the motion.

\subsubsection{Influence of gas density}

For $\bar{\rho}\leq 0.01$, which covers the range of realistic liquid-gas systems, the influence of the finite gas density on the dynamics of coalescence are seen to be negligible.  Once $\bar{\rho}=0.1$ an effect on the bridge front evolution for small radii can be observed, but this only becomes relevant for liquid-liquid systems, which are not considered here in any detail.  Therefore, henceforth the effect of this parameter will not be considered.

\subsubsection{Influence of gas viscosity}\label{S:visc}

\begin{figure}
     \centering
\includegraphics[scale=0.3]{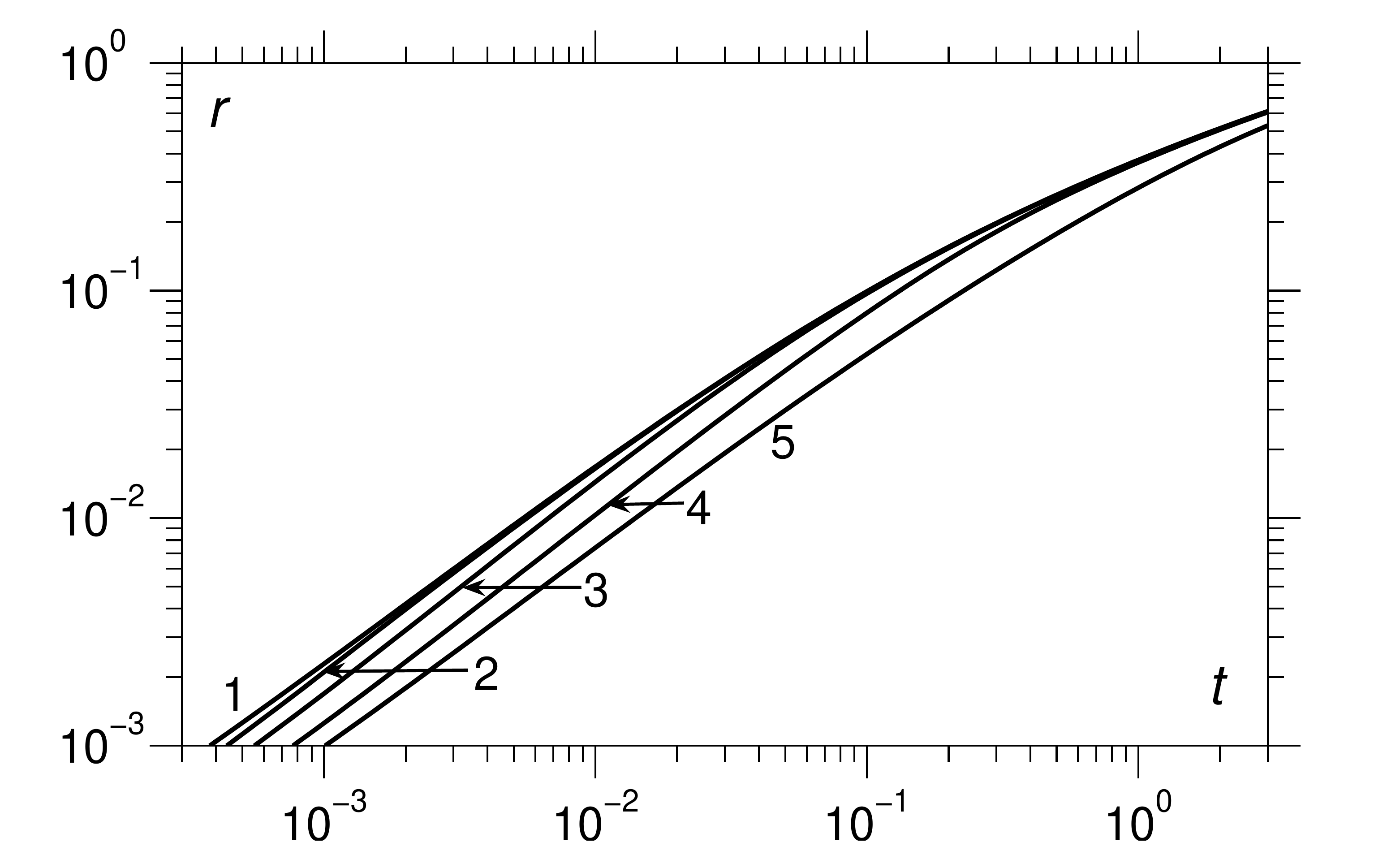}
 \caption{Influence of gas viscosity on the time-dependence of the radius of the bridge connecting the
 coalescing drops calculated for fixed $Re=10^2$ with curve 1: $\bar{\mu}=0$, 2: $\bar{\mu}=10^{-6}$, 3: $\bar{\mu}=10^{-4}$, 4: $\bar{\mu}=10^{-2}$ and 5: $\bar{\mu}=1$.}
 \label{F:mubar}
\end{figure}

Consider now how the viscosity ratio $\bar{\mu}$ affects the coalescence event.  First, taking $Re=10^2$ as a representative case, we show in Figure~\ref{F:mubar} that the viscosity of the gas does have an influence on the initial stages of coalescence and that, as one would hope, for very small viscosity ratio, e.g.\ for $\bar{\mu}=10^{-6}$ (curve 2), the result is almost indistinguishable from the case of a passive gas examined in \S\ref{S:Re} (curve 1).  At the highest viscosity ratio considered $\bar{\mu}=1$ (curve 5), the effect is rather substantial, with a noticeable difference from the passive gas situation (curve 1) well past $r=0.1$. The viscosity ratio of $\bar{\mu}=1$ is, of course, unrealistic for liquid-gas systems, but it is entirely relevant to liquid-liquid ones to which our analysis fully applies. This reduction in the speed of propagation of the bridge's front is due to the additional energy dissipated in the squeezing of fluid out of the thin gap formed ahead of the bridge (Figure~\ref{F:toroidal}).

\begin{figure}
     \centering
\includegraphics[scale=0.3]{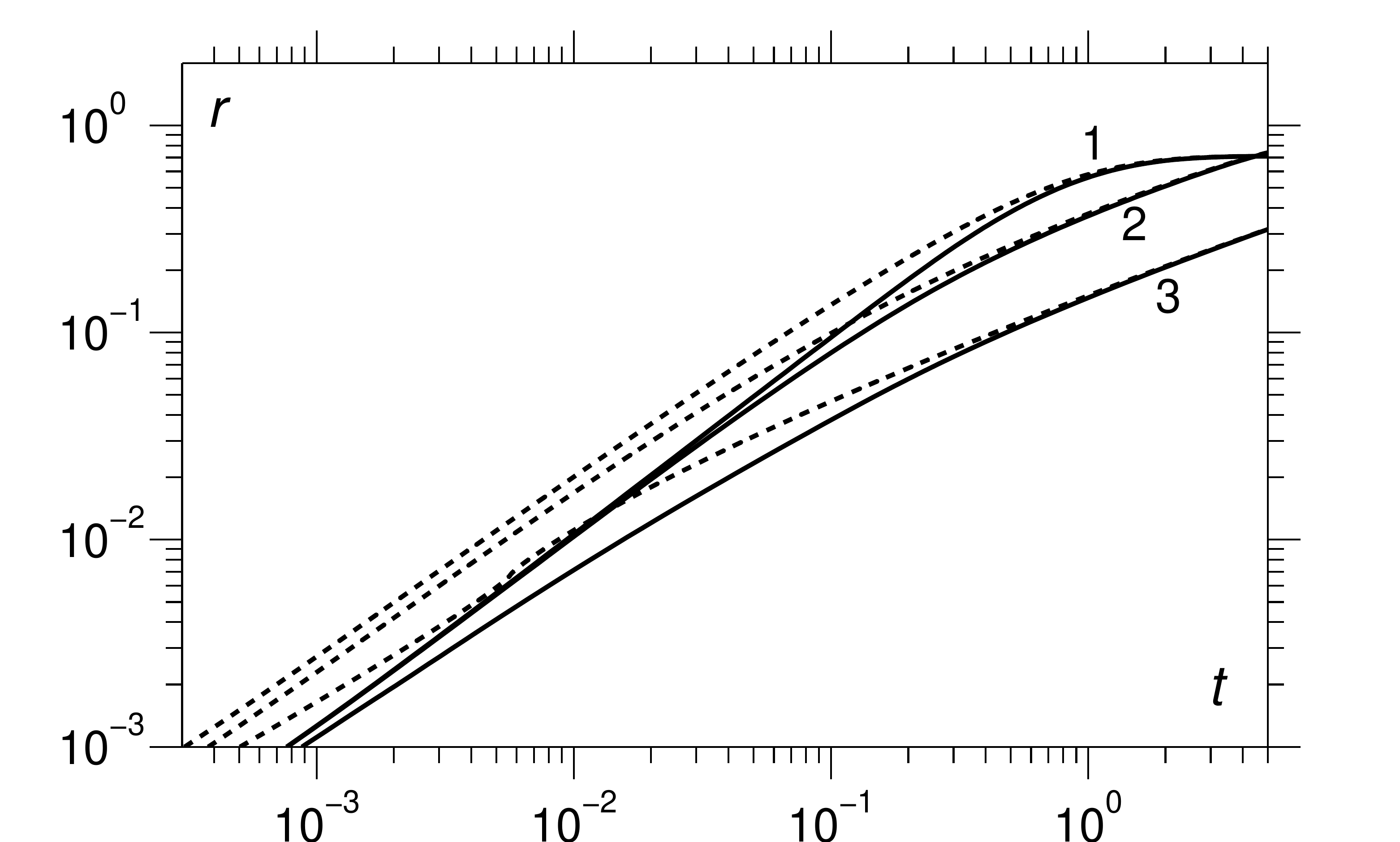}
 \caption{The effect of the Reynolds number on the time-dependence of the radius of the bridge connecting the
 coalescing drops calculated for a fixed viscosity ratio $\bar{\mu}=10^{-2}$ (solid lines) compared to the $\bar{\mu}=0$ case (dashed lines) with curve 1: $Re=0$, curve 2: $Re=10^2$ and curve 3: $Re=10^4$.}
 \label{F:mubar1}
\end{figure}

Figure~\ref{F:mubar1} shows how the inclusion of a gas ($\bar{\mu}=10^{-2}$) affects the coalescence process at different Reynolds numbers (in the liquid, as the inertial effects in the gas have a negligible influence).  In all cases, the gas has a noticeable effect on the motion compared to the passive gas cases (dashed lines) but what is particularly interesting is that for \emph{all}  $Re$ considered, the curve from the viscous-gas case converges to the passive-gas one at around $t=1$, i.e.\ dimensionally at the viscous time scale $T_v = \mu R/\sigma$.  However, at this time ($t=1$), the bridge radii depends on $Re$, with a smaller the Reynolds number (curve 1 is for $Re=0$) giving a larger bridge radius.  In other words, we observe that for a fixed viscosity ratio, the lower the Reynolds number is in the liquid, the more of the coalescence process is affected by the presence of the gas.  This will help to explain our findings in \S\ref{S:exp} where the effects of $Re$ and $\bar{\mu}$ can no longer be varied independently.

\subsubsection{Scaling laws to account for the gas' influence}

In \S\ref{S:Re} it has been shown for the passive gas case that, for $Re=\bar{\mu}=0$, equation (\ref{hopper}) accurately approximates the initial stages of motion whilst (\ref{eggers}) is less useful. However, whilst (\ref{hopper}) is exclusively for one-phase motion without any indication of what effect the dynamics of an ambient gas may have on the initial stages, the scaling law (\ref{eggers}) predicts that taking into consideration the viscosity of the gas will slow the initial stages of coalescence by a factor of four, and, notably, this change in behaviour is predicted to be independent of the viscosity ratio for $r<\bar{\mu}^{2/3}$.

\begin{figure}
     \centering
\includegraphics[scale=0.3]{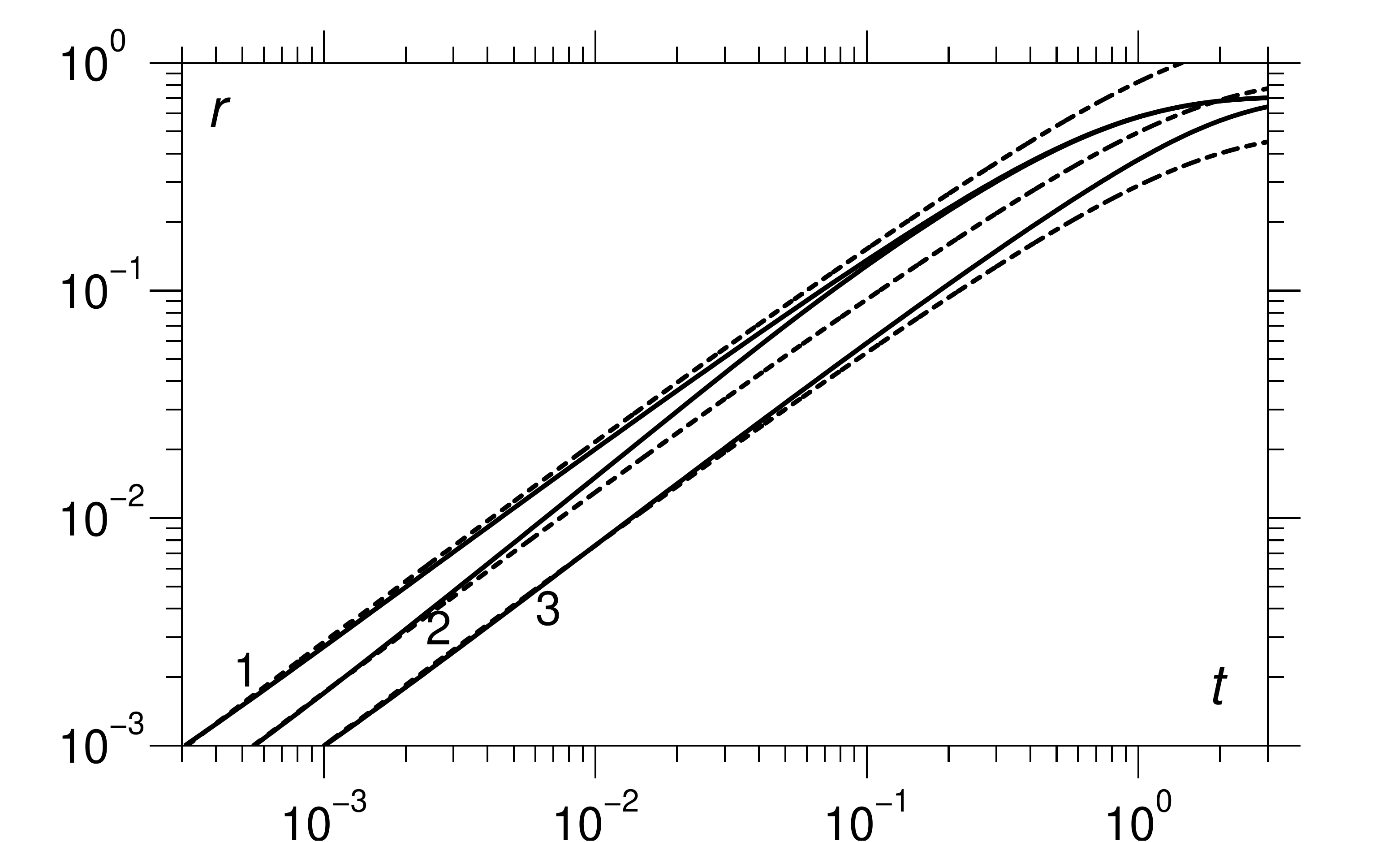}
 \caption{An illustration of the influence of the gas viscosity on the coalescence
 of drops (with $Re=0$). Curve~1: $\bar{\mu}=0$, 2: $\bar{\mu}=10^{-4}$ and
 3: $\bar{\mu}=1$.  The top dashed line correspond to (\ref{hopper}) whilst the middle and lower ones are the same expression with a \emph{fitted} prefactor of $0.6$ and $0.35$, respectively.}
 \label{F:mubareffect}
\end{figure}

From Figure~\ref{F:mubareffect}, where the effect of switching from an inviscid exterior (curve 1) to viscous one (curves 2,~3) at $Re=0$ is considered, we can immediately see that (\ref{eggers}) is both qualitatively and quantitatively incorrect:  the initial stages depend strongly on the viscosity ratio with a larger $\bar{\mu}$ resulting in a slower coalescence. In particular, curve 3, for $\bar{\mu}=1$, is always well below curve 2, obtained for $\bar{\mu}=10^{-4}$.

In an attempt to quantify the effect of the viscosity ratio, the dashed lines in Figure~\ref{F:mubareffect} are equation (\ref{hopper}) with fitted prefactors, i.e.\, instead of computing $r = f(t)$ given exactly by (\ref{hopper}), we consider $r=H f(t)$, where $H$ is a constant chosen to produce a best fit.  It is found that for $\bar{\mu}=0,10^{-4},1$ these prefactors are, respectively, $H=1,0.6,0.35$.  This approach does not originate from any theory, but is simply intended to estimate the effect which a viscous gas has on the motion. It shows, in particular, that for a small viscosity ratio $\bar{\mu}=10^{-4}$, the initial bridge speed is still significantly decreased, but that this effect does not last long, whilst for matched viscosities $\bar{\mu}=1$ the bridge speed is decreased by roughly a factor of three for almost all of the initial stage of coalescence ($r<0.1$). Notably, the different deviations of the dashed lines from the computed solutions suggest that any attempt to somehow make minor adjustments to the passive gas case to account for a viscous exterior are unlikely to work.

\subsubsection{Summary}

The following has been observed for liquid drops coalescing in the viscous gas usually considered in experiments, e.g.\ air at atmospheric pressure, for $Re\in(0,10^4)$:
\begin{itemize}
  \item Inertial effects in the gas have a negligible effect.
  \item The viscosity of the gas prevents toroidal bubble formation.
  \item The bridge evolution can be substantially slower in its initial stages, an effect which increases with the gas-to-liquid ratio $\bar{\mu}$.
  \item For fixed $\bar{\mu}$, lower $Re$ results in the coalescence process being affected on a larger length scale.
  \item Equation (\ref{eggers}) does not capture the aforementioned effects.
\end{itemize}

\section{Comparison to experiments}\label{S:exp}

Having performed a systematic study of the conventional model's predictions, we now proceed to compare these to experimental data where parameters can no longer be independently varied.  In particular, in the experiments of \cite{paulsen11}, the liquids are water-glycerol
mixtures, whose viscosity varies in the range of $\mu=2$--$230$~mPa~s, whilst the density ($\rho=1200$~kg~m${}^{-3}$) and surface
tension with air ($\sigma=65$~mN~m${}^{-1}$) remain approximately the same. These experiments were conducted in air of density
$\rho_g=1.2$~kg~m${}^{-3}$ and viscosity $\mu_g=18$~$\mu$Pa~s. Therefore, as the viscosity is varied, the Reynolds number and viscosity ratio are no longer independent, and we have $\bar{\mu} = 4.6\times10^{-5}Re^{1/2}$. Using these material parameters, we arrive at
\begin{equation}\label{range}
Re\in(1, 10^5),\qquad \bar{\mu} = 4.6\times10^{-5}Re^{1/2}\in(10^{-4},10^{-2}),\qquad \bar{\rho}=10^{-3}.
\end{equation}

Notably, the range in (\ref{range}) has already been covered in the parametric study of \S\ref{S:para}, so that all that remains to be done is to compare the predictions of the conventional model to experimental data.

\subsection{Collapse of data onto a `master curve'}\label{S:master}

In \cite{paulsen11}, it is shown that data for the initial stages of  bridge front evolution, collected from electrical measurements of the coalescence event over a range of different viscosity liquid, can be collapsed onto a master curve:
\begin{equation}\label{collapse}
r/r_c = \frac{2}{1/(t/t_c) + 1/\sqrt{t/t_c}}
\end{equation}
where $t_c,r_c$ are referred to as the (dimensionless here) `crossover' time and radius, where the dominant term in (\ref{collapse}) changes. In particular, for $t\ll t_c$, there is linear growth $r/r_c\sim2t/t_c$, and for later times, $t\gg t_c$, the scaling is of square-root type $r/r_c\sim2\sqrt{t/t_c}$.  Fitting $r_c$ and $t_c$ for every curve enables the dependence of the crossover time on $Re$ to be established.

\begin{figure}
     \centering
\includegraphics[scale=0.3]{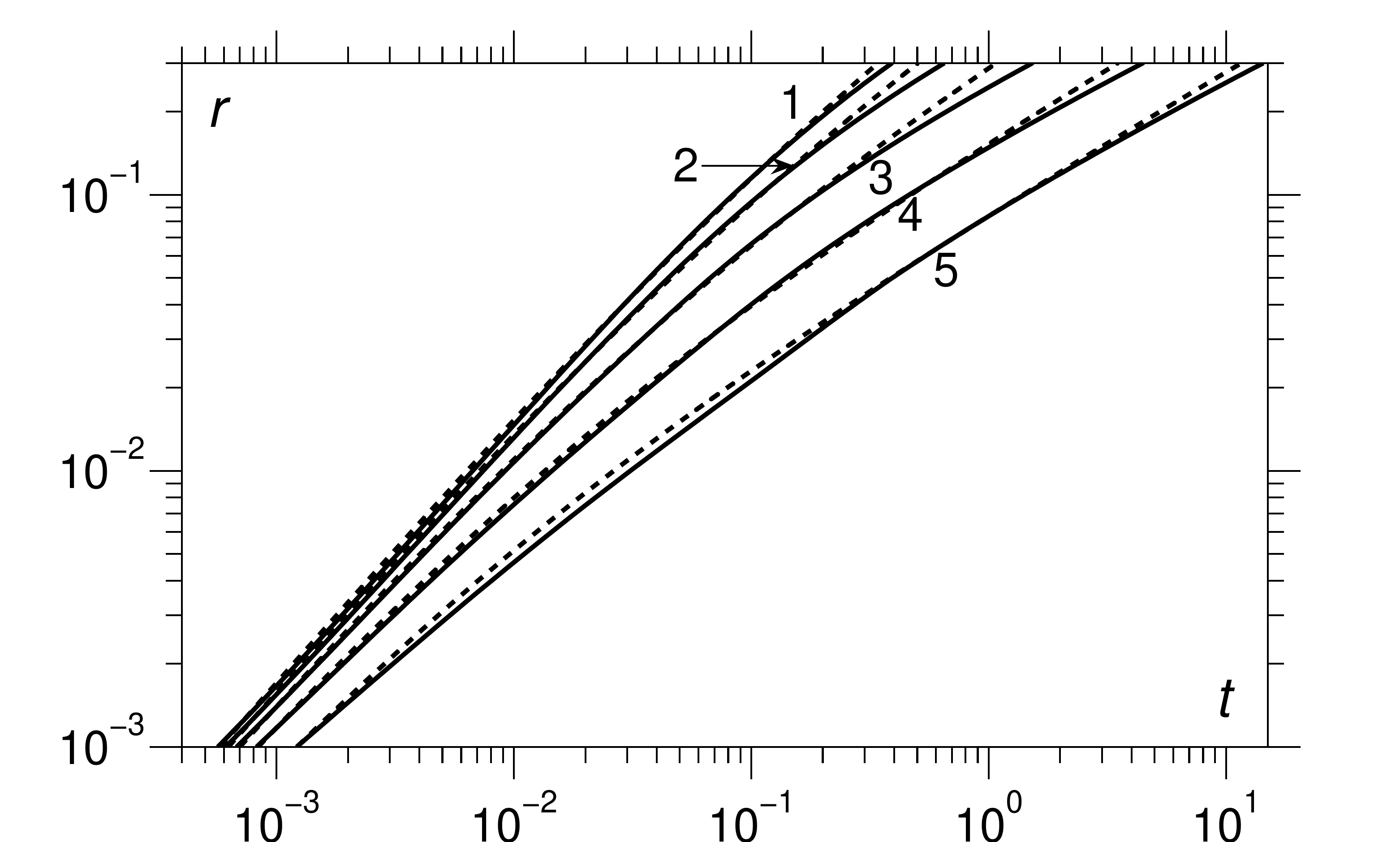}
\includegraphics[scale=0.3]{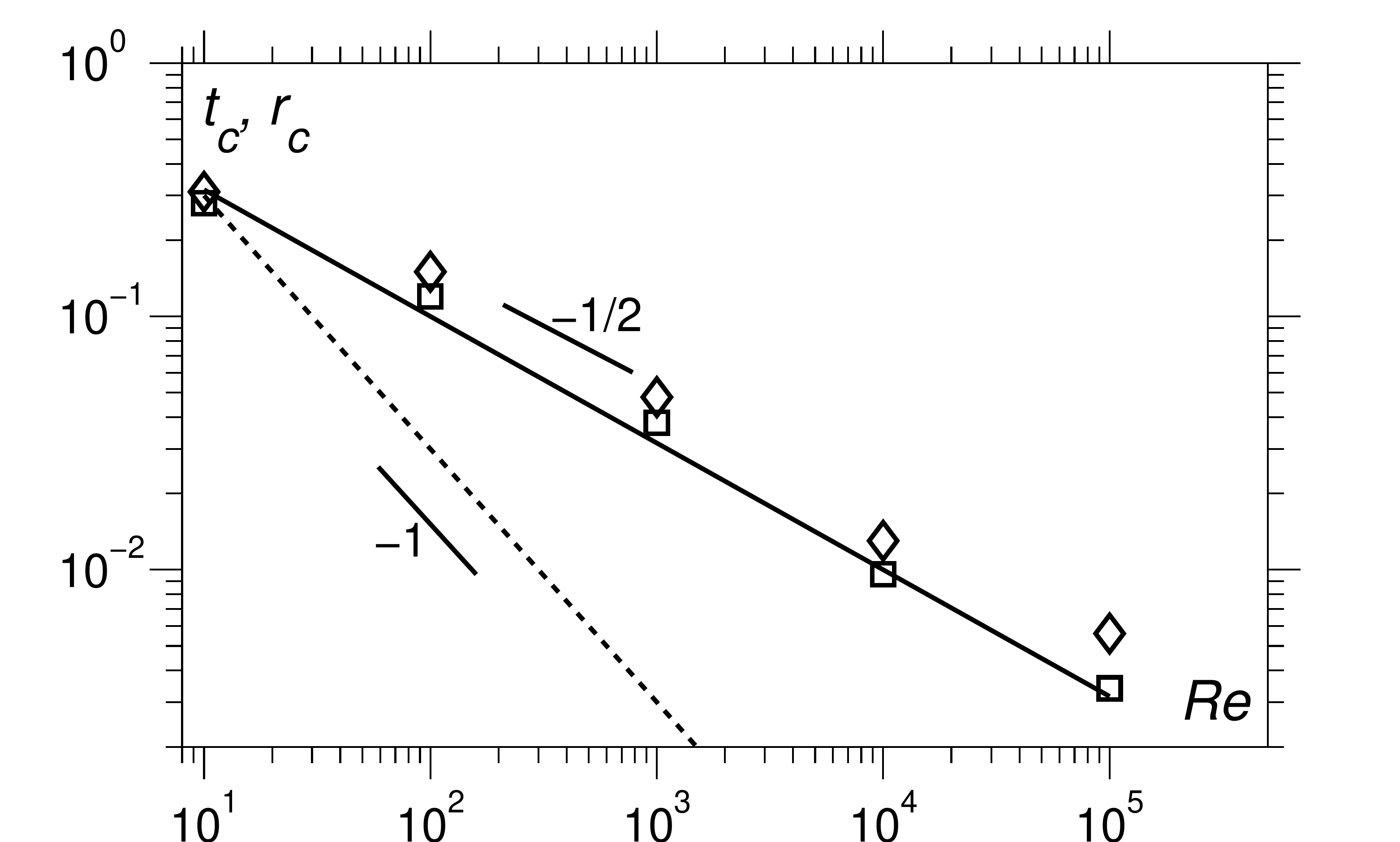}
 \caption{Effect of the $Re$, with $\bar{\mu} = 4.6\times10^{-5}Re^{1/2}$ for the case of a varying viscosity. Curve 1: $Re=10$, curve 2: $Re=10^2$, curve 3: $Re=10^3$, curve 4: $Re=10^4$ and curve 5: $Re=10^5$.  Dashed lines are equation (\ref{collapse}) with constants that are plotted in the lower figure: diamonds are $t_c$ and squares are $r_c$.}
 \label{F:collapse}
\end{figure}

It is of interest to see if a similar fit can be performed with the theoretical curves obtained from the conventional model. In Figure~\ref{F:collapse}, curves 1--5 for $Re=10$--$10^5$ are fitted with the master curve by choosing $(t_c,r_c)$ such that (\ref{collapse}) goes through the computed curves at $r=10^{-3}$ and $r=10^{-1}$.  The fit is relatively good, and the parameters used are plotted in Figure~\ref{F:collapse} as a function of Reynolds number.  Notably, as seen in the experiments \cite{paulsen11}, crossover time and radius scale with $Re^{-1/2}$ as opposed to with $Re^{-1}$, as suggested in some previous works, e.g.\ \cite{eggers99,wu04,aarts05}.  What this essentially means is that the characteristic length scale $L$ appearing in the Reynolds number that determines the crossover value $Re_c$ should be the (dimensional) bridge height $L\sim r_{dim}^2/R$ so that the crossover occurs when $Re_c = \rho \sigma r_{dim}^2/(\mu^2 R) \sim 1$, i.e.\ when $r \sim Re^{-1/2}$.

\subsection{Direct comparison to experimental data}

The analysis in \S\ref{S:master} suggests that many of the trends observed in the experiment are also seen from the computations using the conventional model.  Here, a more direct comparison between theory and experiment, going further than simply confirming the correct scaling behaviour, is performed for the liquids in \cite{paulsen11} with
viscosities $\mu=3.3,~48,~230$~mPa~s as for these mixtures $\sigma$
and $\rho$ vary least ($\rho=1200$~kg~m${}^{-3}$ and
$\sigma=65$~mN~m${}^{-1}$). (The required information about the
mixtures was provided to us by Dr~J.D.~Paulsen, Dr~J.C.~Burton and
Professor S.R.~Nagel.) For the chosen mixtures, one has
$Re=1.4\times10^4,~68,~2.9$. To elucidate the role of the gas'
viscosity, we will look at the difference between the coalescence
occurring in a passive gas studied earlier
\citep{sprittles_pof2} and, as in experiments, in air of density
$\rho_g=1.2$~kg~m${}^{-3}$ and viscosity $\mu_g=18$~$\mu$Pa~s. Then,
the gas-to-liquid density ratio is $\bar{\rho} = 10^{-3}$ and the
viscosity ratios are, respectively, $\bar{\mu} =
5.5\times10^{-3},~3.8\times10^{-4},~7.8\times10^{-5}$.
\begin{figure}
     \centering
\includegraphics[scale=0.3]{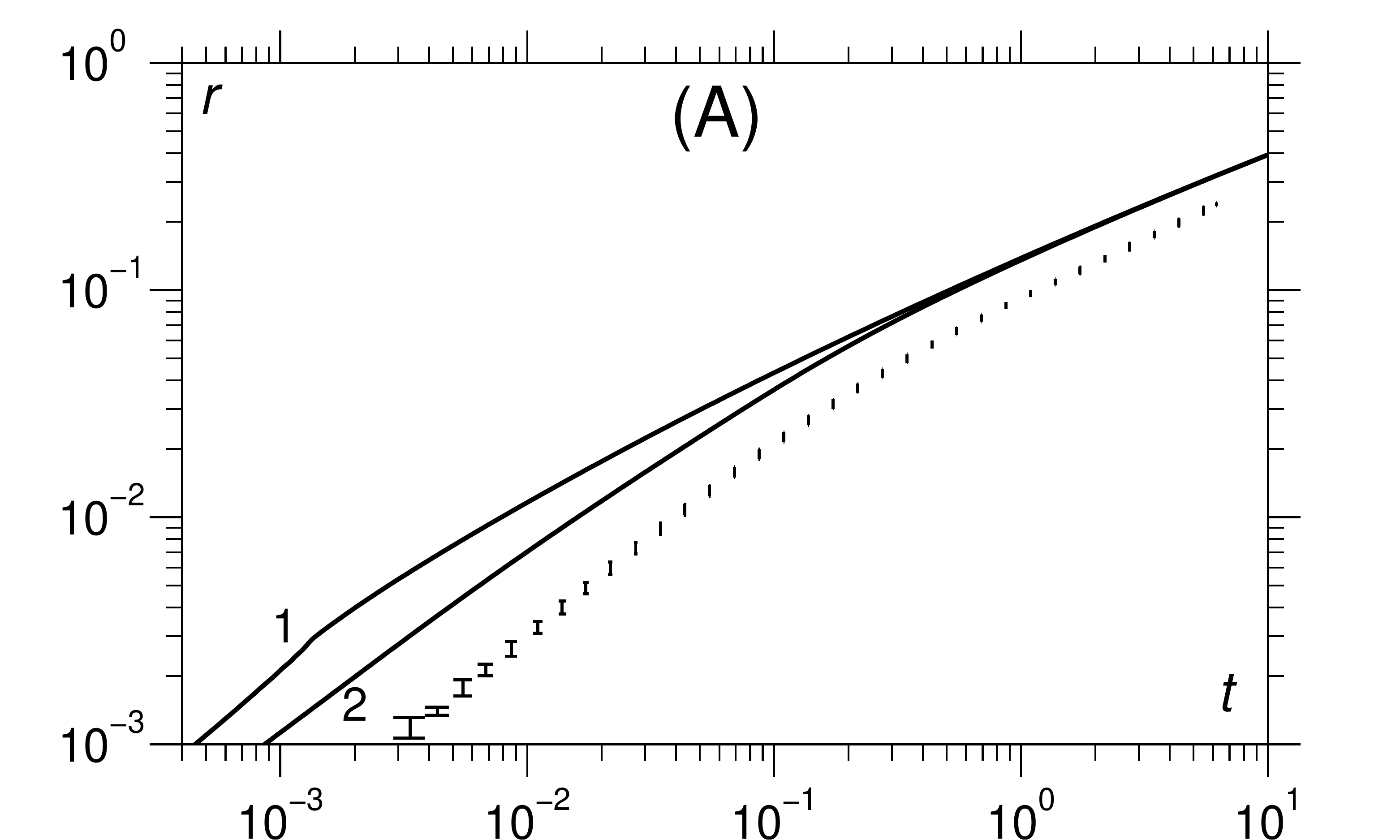}
\includegraphics[scale=0.3]{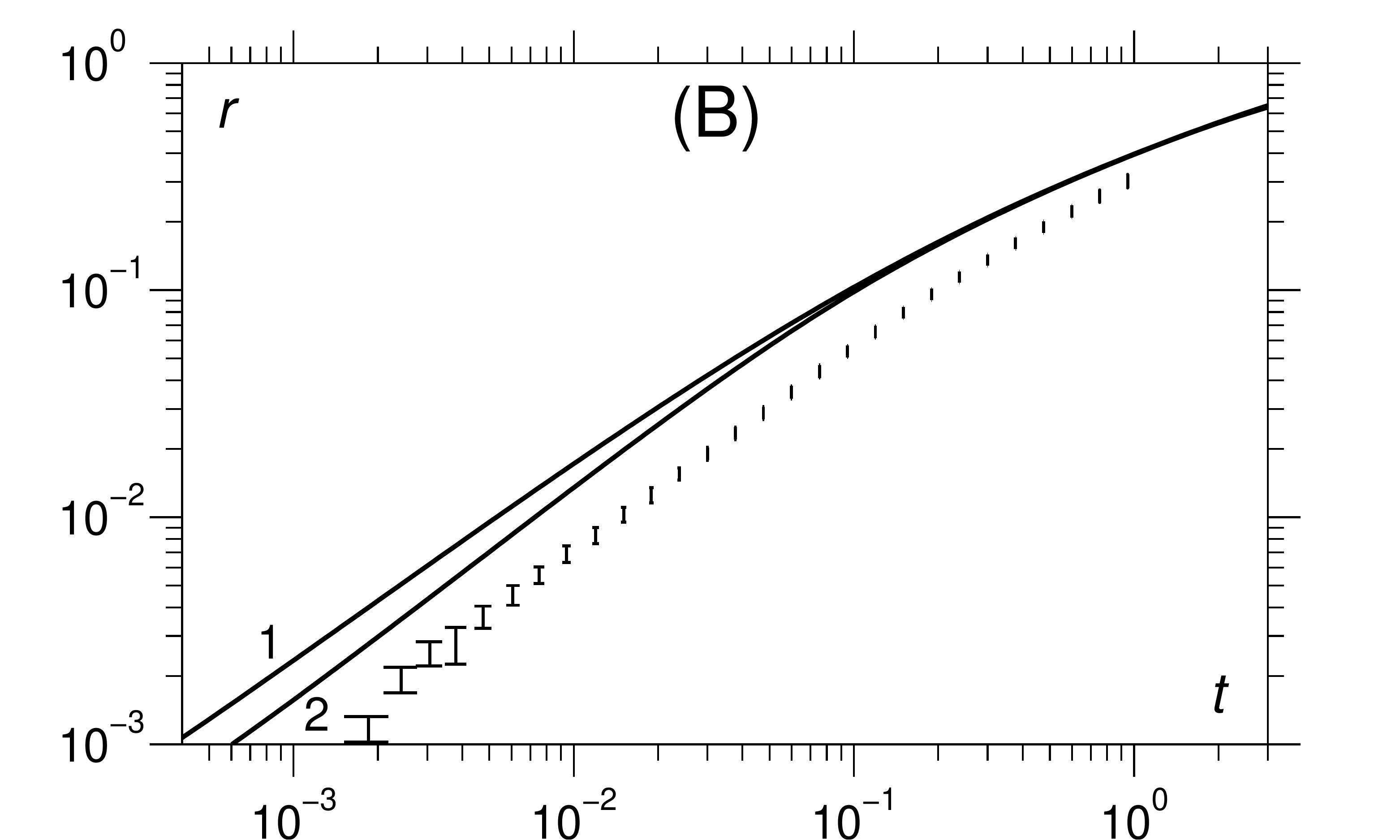}
\includegraphics[scale=0.3]{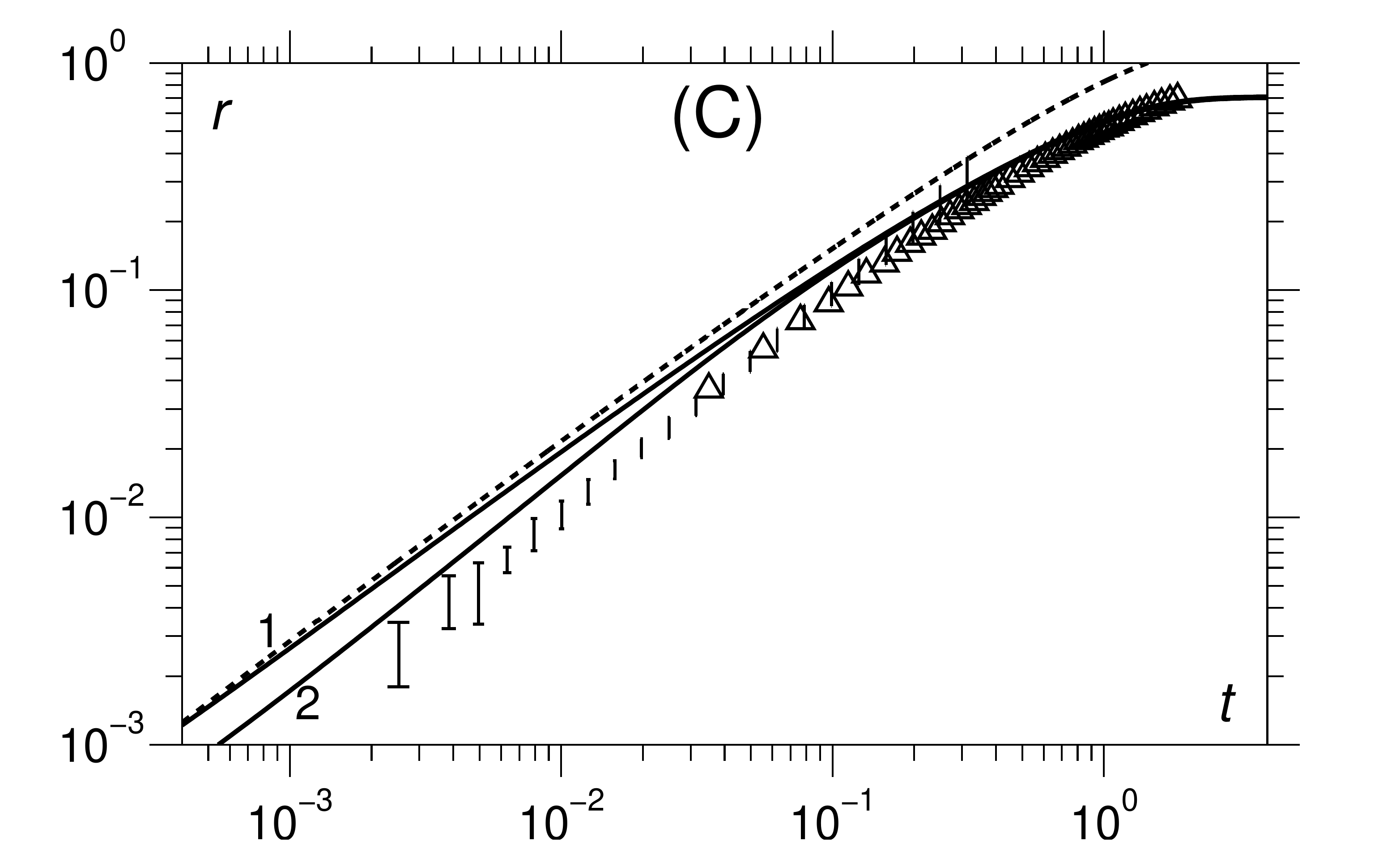}
 \caption{The time-dependence of the radius of the liquid bridge
 connecting the coalescing drops obtained in the framework of the
 conventional model. Curve~1: the drops are in a passive
 gas; curve~2: the ambient gas is viscous with  ($Re,\bar{\mu}$)=($1.4\times10^4,5.5\times10^{-3}$) in (A),
 ($68,3.8\times10^{-4}$) in (B) and ($2.9,~7.8\times10^{-5}$) in (C).  The dashed line in (C) corresponds to (\ref{hopper}).
 The error bars are from experiments in \cite{paulsen11}, and the triangles are
 from optical observations in \cite{thoroddsen05}.}
 \label{F:paulsen}
\end{figure}

Importantly, as one can clearly see in Figure~\ref{F:paulsen}, over the range of viscosities considered, the
presence of the gas does slow down the evolution of the bridge front (curves 2), as compared to the case of a passive gas (curves 1), but this effect is not sufficient to account for the discrepancy
between the conventional model's predictions and the experimental
data from \cite{paulsen11} over the entire period of the experiment.
In particular, although the gas viscosity slows the speed of the
initial motion down, even for the relatively high-viscosity liquid drops, the conventional model still overshoots the
data for the initial stages of the experiment.

It is interesting to see that, roughly, the magnitude of the effect which the introduction of a viscous gas has on the bridge evolution is the same across two orders of magnitude in liquid viscosity.  The reason is that although the viscosity ratio with air decreases with increasing liquid viscosity, the Reynolds number also decreases and, as shown in \S\ref{S:visc}, this results in the gas' influence becoming larger.  These two opposing effects appear to approximately balance each other.

\section{Discussion}\label{S:disc}

Consider now what has been learnt about the initial stages of bridge propagation described in the framework of the conventional model and how this ties in with previously published experimental and computational studies.

\subsection{The presence of an inertially-limited-viscous regime for $Re\leq1$}

It has been shown that for a passive gas, the Stokes flow solution (\ref{hopper}) describes the initial stages of growth for $Re\leq1$, a result that is in direct conflict with the conclusions of \cite{paulsen12} which claim that the Stokes flow solution is only entered after the ILV regime has occurred.  In \cite{paulsen12}, a key observation in favour of the ILV regime is that, at finite $Re$, it takes a certain time for the apex of the drop to follow the Stokes flow solution, and we have also observed this phenomenon.  How then, do these apparently contradictory findings square with each other?

First, the computations performed in \cite{paulsen12}, see for example Fig.~3E there, are for $r>10^{-2}$ which only leaves the interval $10^{-2}<r<10^{-1}$ to consider the initial stage of motion.  Consequently, the good agreement of Hopper's solution (\ref{hopper}) with full computations at $Re\sim1$, for $r\ll1$, confirmed here in Figure~\ref{F:paulsen} for exactly the same case as the one in Fig.~3E of \cite{paulsen12}, appears to have been missed.  Instead, in \cite{paulsen12}, the results of the computations are shown to give an approximately linear growth in the bridge radius and this is used as evidence against the Stokes regime.  We have seen that this is not the case.  For $Re\leq 1$, $\bar{\mu}=0$ and $r\ll1$, the bridge propagation is best described by Hopper's solution (\ref{hopper}) corresponding to the Stokes regime.

Far from the bridge front, at finite $Re$, the inertia is important as the flow then takes some time to develop, in contrast to the $Re=0$ case.  Thus, if one considers the \emph{global} motion of the drops for $Re\leq1$, it makes sense to talk of an ILV regime, even though \emph{local} to the bridge front the finiteness of $Re$ has a negligible effect.

Second, we would like to tie these observations in with the experimental findings in \cite{paulsen11,paulsen12}.  This is quite tricky, as it involves considering the effect of the gas on the motion as well as recognising that quantitatively the experiments do not agree well with the predictions of the conventional model (Figure~\ref{F:paulsen}).  One thing that can be noted; however, is that there has been no systematic experimental investigation of the regime $Re\leq1$ and $r\ll1$.  In \cite{paulsen11} the electrical measurements allowed for $r\ll 1$, but all data was for $Re\geq 1$, whilst in \cite{paulsen12} hanging pendent drops of huge viscosity were considered, so that $Re\leq 1$, but only optical measurements were made, so that $r>0.1$.  Further experiments on this regime may reveal more details about the initial stages of motion.

\subsection{Characterising parameter space}

Much has been made about the different `regimes' of coalescence, their different `scalings' and the possibility of collapsing all data onto a master curve using two fitting parameters. However, let us consider instead the question of when such simplified models actually allow us to ascertain accurate quantitative information about the coalescence event. First, such data about the entire drop shape is impossible, as it is only in the two-dimensional case that the theory of \cite{hopper84} applies, and the other works all consider only predictions for the bridge radius as a function of time, i.e.\ `local' information.  Moreover, to increase our chances of progress in this task, let us further simplify matters by considering the gas to be passive, so that the only governing parameter is then the Reynolds number. So the question essentially becomes, at a given $Re$, at what bridge radii $r$ is there a quantitative formula which relates $r$ to time $t$?

This question has been addressed in previous works, e.g.\ \cite{paulsen12}, but the difference between their approach and the one we take here is that we are interested in where \emph{quantitative predictions} can be made, rather than where \emph{qualitative behaviour} occurs.  Mathematically, the difference is that, whilst previous works have put all coalescence events where the bridge radius scales in a certain way, e.g.\ linear $r=C_v t$, into one regime, with $C_v$ fitted to the data in an arbitrary way, here, we will only consider regimes in which there are no fitted prefactors, e.g.\ (\ref{hopper}), or those in which the prefactor is known and fixed.  This is not a better approach than \cite{paulsen12}, it is just a different one, motivated by a desire to understand in which parts of parameter space quantitative predictions using simple analytic formulas can be made.

To be precise, consider the error $E(t)$ between a computed solution $r(t)$ and an approximate expression $r_{approx}(t)$ to be given by
\begin{equation}\label{error}
E = \frac{|r-r_{approx}|}{r}
\end{equation}
and consider, for a given $Re$, the values of the radius $r$ for which relative error falls below $10$\% ($E<0.1$), i.e.\, very crudely, below experimental error.  The result will be that for each $r_{approx}$ there is a section of $(Re,r)$ phase space in which the approximate expression meets the required tolerance. As with the computations, only $r>10^{-3}$ is considered, as before this point there are large relative errors associated with the finite initial bridge radius from which the computations start.

\begin{figure}
     \centering
\includegraphics[scale=0.3]{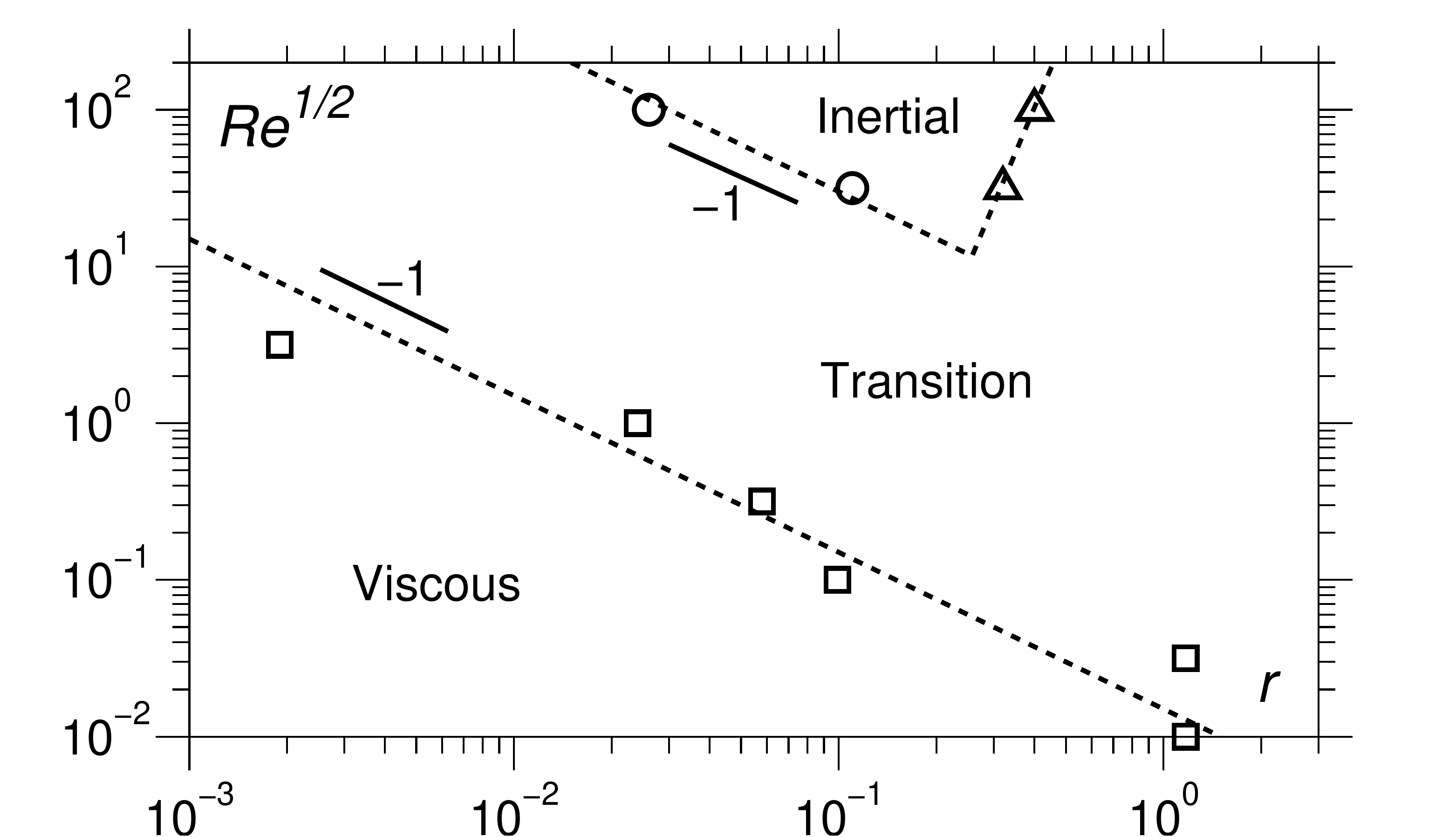}
 \caption{Phase diagram showing the regions in which the bridge radius $r$ is in the viscous regime described by (\ref{hopper}) or in the inertial regime described by (\ref{io}) with $C_i=1.5$.  Dashed lines are rough fits to the data with the lower dashed line given by $Re^{1/2}=1.5\times 10^{-2}r^{-1}$, upper dashed line (with gradient $-1$) is $Re^{1/2}=3r^{-1}$ and the third dashed line, going through the triangles, is simply to guide the eye.}
 \label{F:phase}
\end{figure}

In Figure~\ref{F:phase}, the new phase diagram is shown which, to be consistent with previous works, has been produced for the case of free spheres coalescing.  A phase diagram for pinned hemispheres differs very little, as it is only in the later stages of motion, $r>0.1$, that geometry starts to have an effect.  It makes sense to plot $r$ against $Re^{1/2}$ rather than $Re$ as (a) we see that the boundaries to the different regions of phase space are given by $r\propto Re^{-1/2}$ and (b) by flipping the curves about the plane $Re=1$, the plot becomes $r$ vs $Oh=Re^{-1/2}$ so that a comparison to the phase diagram in \cite{paulsen12} can more easily be performed.

Square markers show the region in which Hopper's solution (\ref{hopper}) accurately describes the computed solutions to within the required tolerance.  These quantitative results confirm that, in the initial stages of motion, the Stokes flow solution describes the bridge's dynamics for a distance $r=r(Re)$ and that this distance scales like $Re^{-1/2}$.  Exiting the viscous regime does not mean that the motion is then in an inertial regime.  In actual fact, there is a large part of parameter space where the motion can neither be considered viscosity-dominated nor inertia-dominated.  Although it is tempting to call this the inertially-limited viscous regime, and in some sense it is, as it is the region where inertial and viscous forces are important, we have labelled this region `Transition' as, to be consistent with our aims, there is no quantitative predictive expression for this region, and all what we know is that the process goes across this regime from `viscous' to `inertial'.

For high enough $Re$, roughly $Re>100$, an inertial regime is entered in which (\ref{io}) with $C_i=1.5$ accurately describes the computed solution.  Notably, this region has both a lower bound, as it takes some distance for viscous effects to become negligible, and an upper bound at which point the assumptions made in (\ref{io}), such as the motion being driven entirely by the longitudinal curvature, no longer hold, see \cite{sprittles14_pre} for further details.
%

The phase diagram we have created can be very simply interpreted.  In the viscous region, the motion can be approximated by the Stokes equations, i.e.\ neglecting the inertial terms, whilst in the inertial regime the viscous terms are negligible, so that the Euler equations should be able to describe the motion. Elsewhere the full Navier-Stokes equations are necessary for accurate computation and thus simplified expressions based on the aforementioned limiting cases cannot in principle be accurate.

The computed two-dimensional phase diagram is actually a cross-section ($\bar{\mu}=0$) of the three-dimensional parameter space $(r,\sqrt{Re},\bar{\mu})$ which would be required if the viscosity ratio was also accounted for. At moderate $\bar{\mu}$, it is likely that Hopper's solution will no longer become an accurate representation of the initial stages, so that no currently-available quantitative expressions exist for this period. It may be that in this case, a linear expression, as proposed in \cite{paulsen11}, describes the data well, but that the required prefactor's dependency on $Re$ and $\bar{\mu}$ will be a-priori unknown.  Thus, in this situation, the region in which computations are required to provide quantitative predictions of the coalescence phenomenon will inevitably grow.

\section{Outlook}\label{S:conc}

Using computational techniques, a systematic parametric study of the process of coalescence in the framework of the conventional model has been performed and has enabled us to identify a number of misconceptions in the published literature and suggest avenues of further research.

In particular, our results have shown that:
\begin{enumerate}
  \item When viscous forces dominate inertial ones, Hopper's solution \citep{hopper84} best approximates the initial stages of coalescence \emph{local} to the bridge front and the inertially-limited viscous regime is seen to be a characteristic of the \emph{global} motion of the drops. In contrast, experimental results in \cite{paulsen12,paulsen13,paulsen14} indicate that this regime also affects the \emph{local} motion of the bridge front.  The reason for this discrepancy remains unexplained.
  \item There is a `transition region' in which, currently, there is no predictive analytic theory.
  \item Toroidal bubbles are not formed for coalescence of liquid drops in air at atmospheric pressure.
  \item The conventional model captures the scaling behaviour of the transitions between different regimes observed in experiments, but quantitatively overshoots the data for $r$ vs $t$.
\end{enumerate}

Each of these findings suggests a particular avenue of enquiry deserving of further attention:
\begin{enumerate}
  \item Electrical methods focused on the very initial stages of coalescence for high-viscosity liquids (low $Re$) would determine whether or not experimental measurements agree with the conventional model's prediction that this regime can be described by Hopper's solution.
  \item If it is possible to develop an asymptotic theory for the transition regime, that gives a simplified framework into which the predictions of the conventional model can be understood, much like Hopper's solution for the viscous-regime, then this should be considered and our results would provide a benchmark for it.  If not, as seems likely, particularly when considering the influence of an ambient fluid as well, then computational techniques should be recognised as the only approach giving quantitative predictions for this regime.
  \item To attempt to reach the regime in which toroidal bubble formation can be observed, one must consider lowering the influence of the gas viscosity.  This could potentially be realised by reducing the ambient pressure of the gas. Simulations in this regime may shed further light on this possibility and thus aid any experimental attempts.
  \item Perhaps most importantly, experimental and theoretical aspects of the coalescence process should be reconsidered in light of the poor quantitative agreement between electrical measurements and the predictions of the conventional model.  Two possibilities for the discrepancy are that (i) there is an effect in the experiment which is not accounted for in the theory, such as the influence of the electric field on the motion or (ii) that the conventional model itself is unable to capture the initial stages of motion due to its singular nature, and, if this is the case, then singularity-free descriptions that incorporate extra physics, such as the interface formation model considered in \cite{sprittles_pof2,sprittles14_jfm1}, deserve further attention.
\end{enumerate}
%



\section*{Acknowledgements}

The authors would like to thank Dr J.D.~Paulsen, Dr J.C.~Burton and Professor S.R.~Nagel for providing us with the data from their experiments published in \cite{paulsen11,paulsen12}.

\bibliographystyle{jfm}
\bibliography{Bibliography}

\begin{thebibliography}{28}
\expandafter\ifx\csname natexlab\endcsname\relax\def\natexlab#1{#1}\fi

\bibitem[Aarts {\em et~al.\/}(2005)Aarts, Lekkerkerker, Guo, Wegdam \&
  Bonn]{aarts05}
{\sc Aarts, D. G. A.~L., Lekkerkerker, H. N.~W., Guo, H., Wegdam, G.~H. \&
  Bonn, D.} 2005 Hydrodynamics of droplet coalescence. {\em Physical Review
  Letters\/} {\bf 95}, 164503.

\bibitem[Derby(2010)]{derby10}
{\sc Derby, B.} 2010 Inkjet printing of functional and structural materials:
  fluid property requirements, feature stability and resolution. {\em Annual
  Review of Materials Research\/} {\bf 40}, 395--414.

\bibitem[Duchemin {\em et~al.\/}(2003)Duchemin, Eggers \&
  Josserand]{duchemin03}
{\sc Duchemin, L., Eggers, J. \& Josserand, C.} 2003 Inviscid coalescence of
  drops. {\em Journal of Fluid Mechanics\/} {\bf 487}, 167--178.

\bibitem[Eddi {\em et~al.\/}(2013)Eddi, Winkels \& Snoeijer]{eddi13}
{\sc Eddi, A., Winkels, K.~G. \& Snoeijer, J.~H.} 2013 Short time dynamics of
  viscous drop spreading. {\em Physics of Fluids\/} {\bf 25}, 013102.

\bibitem[Eggers {\em et~al.\/}(1999)Eggers, Lister \& Stone]{eggers99}
{\sc Eggers, J., Lister, J.~R. \& Stone, H.~A.} 1999 Coalescence of liquid
  drops. {\em Journal of Fluid Mechanics\/} {\bf 401}, 293--310.

\bibitem[Enright {\em et~al.\/}(2012)Enright, Miljkovic, {Al-Obeidi}, Thompson
  \& Wang]{enright12}
{\sc Enright, R., Miljkovic, N., {Al-Obeidi}, A., Thompson, C.~V. \& Wang,
  E.~N.} 2012 Condensation on superhydrophobic surfaces: the role of local
  energy barriers and structure length scale. {\em Langmuir\/} {\bf 28},
  14424–--14432.

\bibitem[Hopper(1984)]{hopper84}
{\sc Hopper, R.~W.} 1984 Coalescence of two equal cylinders: exact results for
  creeping viscous plane flow driven by capillarity. {\em Journal of the
  American Ceramic Society\/} {\bf 67}, 262--264.

\bibitem[Hopper(1990)]{hopper90}
{\sc Hopper, R.~W.} 1990 Plane {Stokes} flow driven by capillarity on a free
  surface. {\em Journal of Fluid Mechanics\/} {\bf 213}, 349--375.

\bibitem[Hopper(1993{\natexlab{{\em a\/}}})]{hopper93a}
{\sc Hopper, R.~W.} 1993{\natexlab{{\em a\/}}} Coalescence of two viscous
  cylinders by capillarity: {Part 1}. {Theory}. {\em Journal of the American
  Ceramic Society\/} {\bf 76}, 2947--2952.

\bibitem[Hopper(1993{\natexlab{{\em b\/}}})]{hopper93b}
{\sc Hopper, R.~W.} 1993{\natexlab{{\em b\/}}} Coalescence of two viscous
  cylinders by capillarity: {Part 2}. {Shape evolution}. {\em Journal of the
  American Ceramic Society\/} {\bf 76}, 2953--2960.

\bibitem[Menchaca-Rocha {\em et~al.\/}(2001)Menchaca-Rocha,
  Mart\'{\i}nez-D\'{a}valos, N\'{u}\'{n}ez, Popinet \&
  Zaleski]{menchacarocha01}
{\sc Menchaca-Rocha, A., Mart\'{\i}nez-D\'{a}valos, A., N\'{u}\'{n}ez, R.,
  Popinet, S. \& Zaleski, S.} 2001 Coalescence of liquid drops by surface
  tension. {\em Physical Review E\/} {\bf 63}, 046309.

\bibitem[Oguz \& Prosperetti(1989)]{oguz89}
{\sc Oguz, H.~N. \& Prosperetti, A.} 1989 Surface-tension effects in the
  contact of liquid surfaces. {\em Journal of Fluid Mechanics\/} {\bf 203},
  149--171.

\bibitem[Paulsen(2013)]{paulsen13}
{\sc Paulsen, J.~D.} 2013 Approach and coalescence of liquid drops in air. {\em
  Physical Review E\/} {\bf 88}, 063010.

\bibitem[Paulsen {\em et~al.\/}(2011)Paulsen, Burton \& Nagel]{paulsen11}
{\sc Paulsen, J.~D., Burton, J.~C. \& Nagel, S.~R.} 2011 Viscous to inertial
  crossover in liquid drop coalescence. {\em Physical Review Letters\/} {\bf
  106}, 114501.

\bibitem[Paulsen {\em et~al.\/}(2012)Paulsen, Burton, Nagel, Appathurai, Harris
  \& Basaran]{paulsen12}
{\sc Paulsen, J.~D., Burton, J.~C., Nagel, S.~R., Appathurai, S., Harris, M.~T.
  \& Basaran, O.} 2012 The inexorable resistance of inertia determines the
  initial regime of drop coalescence. {\em Proceedings of the National Academy
  of Science\/} {\bf 109}, 6857--6861.

\bibitem[Paulsen {\em et~al.\/}(2014)Paulsen, Carmigniani, Kannan, Burton \&
  Nagel]{paulsen14}
{\sc Paulsen, J.~D., Carmigniani, R., Kannan, A., Burton, J.~C. \& Nagel,
  S.~R.} 2014 Coalescence of bubbles and drops in an outer fluid. {\em Nature
  Communications\/} {\bf 5}, 3182.

\bibitem[Richardson(1992)]{richardson92}
{\sc Richardson, S.} 1992 Two-dimensional slow viscous flows with
  time-dependent free boundaries driven by surface tension. {\em European
  Journal of Applied Mathematics\/} {\bf 3}, 193--207.

\bibitem[Shikhmurzaev(2007)]{shik07}
{\sc Shikhmurzaev, Y.~D.} 2007 {\em Capillary Flows with Forming Interfaces\/}.
  Chapman \& Hall/CRC, Boca Raton.

\bibitem[Sprittles \& Shikhmurzaev(2012{\natexlab{{\em a\/}}})]{sprittles_pof2}
{\sc Sprittles, J.~E. \& Shikhmurzaev, Y.~D.} 2012{\natexlab{{\em a\/}}}
  Coalescence of liquid drops: different models versus experiment. {\em Physics
  of Fluids\/} {\bf 24}, 122105.

\bibitem[Sprittles \& Shikhmurzaev(2012{\natexlab{{\em b\/}}})]{sprittles_pof}
{\sc Sprittles, J.~E. \& Shikhmurzaev, Y.~D.} 2012{\natexlab{{\em b\/}}} The
  dynamics of liquid drops and their interaction with solids of varying
  wettabilities. {\em Physics of Fluids\/} {\bf 24}, 082001.

\bibitem[Sprittles \& Shikhmurzaev(2012{\natexlab{{\em
  c\/}}})]{sprittles_ijnmf}
{\sc Sprittles, J.~E. \& Shikhmurzaev, Y.~D.} 2012{\natexlab{{\em c\/}}} A
  finite element framework for describing dynamic wetting phenomena. {\em
  International Journal for Numerical Methods in Fluids\/} {\bf 68},
  1257--1298.

\bibitem[Sprittles \& Shikhmurzaev(2013)]{sprittles_jcp}
{\sc Sprittles, J.~E. \& Shikhmurzaev, Y.~D.} 2013 Finite element simulation of
  dynamic wetting flows as an interface formation process. {\em Journal of
  Computational Physics\/} {\bf 233}, 34--65.

\bibitem[Sprittles \& Shikhmurzaev(2014{\natexlab{{\em
  a\/}}})]{sprittles14_jfm1}
{\sc Sprittles, J.~E. \& Shikhmurzaev, Y.~D.} 2014{\natexlab{{\em a\/}}} The
  coalescence of liquid drops in a viscous fluid: interface formation model.
  Journal of Fluid Mechanics: In Press.

\bibitem[Sprittles \& Shikhmurzaev(2014{\natexlab{{\em
  b\/}}})]{sprittles14_pre}
{\sc Sprittles, J.~E. \& Shikhmurzaev, Y.~D.} 2014{\natexlab{{\em b\/}}}
  Dynamics of liquid drops coalescing in the inertial regime. {\em Physical
  Review E\/} {\bf 89}, 063006.

\bibitem[Thoroddsen {\em et~al.\/}(2008)Thoroddsen, Etoh \&
  Takehara]{thoroddsen08}
{\sc Thoroddsen, S.~T., Etoh, T.~G. \& Takehara, K.} 2008 High-speed imaging of
  drops and bubbles. {\em Annual Review of Fluid Mechanics\/} {\bf 40},
  257--285.

\bibitem[Thoroddsen \& Takehara(2000)]{thoroddsen00}
{\sc Thoroddsen, S.~T. \& Takehara, K.} 2000 The coalescence cascade of a drop.
  {\em Physics of Fluids\/} {\bf 12}, 1265--1267.

\bibitem[Thoroddsen {\em et~al.\/}(2005)Thoroddsen, Takehara \&
  Etoh]{thoroddsen05}
{\sc Thoroddsen, S.~T., Takehara, K. \& Etoh, T.~G.} 2005 The coalescence speed
  of a pendent and sessile drop. {\em Journal of Fluid Mechanics\/} {\bf 527},
  85--114.

\bibitem[Wu {\em et~al.\/}(2004)Wu, Cubaud \& Ho]{wu04}
{\sc Wu, M., Cubaud, T. \& Ho, C.} 2004 Scaling law in liquid drop coalescence
  driven by surface tension. {\em Physics of Fluids\/} {\bf 16}, 51--54.

\end{thebibliography}

\end{document}